\documentclass{article}

\usepackage{arxiv}
\usepackage[utf8]{inputenc} 
\usepackage[T1]{fontenc}    
\usepackage{microtype}      
\usepackage{booktabs}       
\usepackage{amsmath}
\usepackage{amsfonts}
\usepackage{amssymb}
\usepackage{nicefrac}       
\usepackage{url}            
\usepackage{hyperref}       
\usepackage{cleveref}       
\usepackage{graphicx}
\usepackage{doi}
\usepackage{mhchem}
\usepackage{stmaryrd}
\usepackage{adjustbox}
\usepackage{esint}
\usepackage{bbold}
\usepackage{subcaption}
\usepackage{float}
\usepackage{mathrsfs}
\usepackage{soul}
\usepackage{xcolor}
\usepackage{tcolorbox}
\usepackage{tikz}
\usepackage{placeins}
\usepackage{appendix}
\usepackage[authoryear]{natbib}

\usepackage{geometry}
\usepackage{multirow}
\usepackage{diagbox}
\usepackage{arydshln}

\graphicspath{{./figures/}}

\title{Detonation propagation in three-dimensional continuous curved ducts}

\usepackage[affil-it]{authblk} 

\begin{document}
\author[1]{Lisong Shi\thanks{ls.mark.shi@gmail.com}}
\author[1]{Chih-Yung Wen\thanks{chihyung.wen@polyu.edu.hk}}
\author[2,3]{Xuxu Sun\thanks{xuxusun@whut.edu.cn}}
\author[1,4]{E Fan\thanks{Corresponding author: e.fan@connect.polyu.hk}}

\affil[1]{Department of Aeronautical and Aviation Engineering, The Hong Kong Polytechnic University}
\affil[2]{School of Safety Science and Emergency Management, Wuhan University of Technology}
\affil[3]{Hubei Key Laboratory of Fuel Cell, Wuhan University of Technology}
\affil[4]{Department of Mechanics and Aerospace Engineering, Southern University of Science and Technology}

\maketitle

\begin{abstract}
In this paper, three-dimensional (3D) detonation numerical studies are conducted using time-dependent reactive Euler equations in both straight and curved channels. Simulations are performed in continuous curved ducts to ensure the rotating detonation waves are fully developed. These simulations are compared to investigate the response of detonation waves to curvature within infinitely long square ducts. The influence of the inner wall radius, cross-section size, and activation energy ($E_{\text {a}}$) on wave structures, pressure distributions, and velocity evolution are carefully described. Additionally, a comparison with two-dimensional (2D) simulations is also provided. The results for detonation waves with low $E_{\mathrm{a}}$ in narrow ducts show that, in straight ducts, it typically exhibits rectangular or diagonal modes near the Chapman-Jouguet speed, and the mode depends on the initial perturbations. However, when propagating in curved ducts, the waves display significantly different patterns and curvature sensitive velocity deficits. For sufficient small radii, due to the compression and expansion in the lateral direction, an initial diagonal perturbation may still quickly transit into rectangular mode. Nevertheless, the soot foils from 2D simulations closely resemble those from 3D simulations with rectangular mode, except for the absence of slapping waves. For detonation waves with low $E_{\mathrm{a}}$ in wide ducts, we find that mode transition may happen even for rectangular perturbations. An out-of-phase rectangular mode first appears, followed by the twisting of the transverse waves until a diagonal mode develops. The corresponding curved case shows that only one pair of transverse waves on each wall. Furthermore, the cellular patterns become irregular with increasing $E_{\text {a}}$. In the straight duct with high $E_{\text {a}}$, the cells seem more randomly distributed. In contrast for curved duct with high $E_{\mathrm{a}}$, small cells are observed on the outer wall, while large-scale wave motions are noted on the inner wall, as a result of mixture with high $E_{\text {a}}$ is more sensitive to the perturbation such as variation in local propagation velocity. Nonetheless, the present results indicate that for a given inner radius and duct size, the propagation velocity are almost independent with the sensitivity of the mixture. This study characterizes, for a fully developed detonation wave in a continuously curved duct, the effect of substantial compression and velocity deficit in altering wave structures along the inner and outer walls. Additionally, it shows that, the number of transverse waves in the horizontal direction (parallel to the upper and bottom walls) can be suppressed.

\end{abstract}

\keywords{three-dimensional detonation \and curved duct \and cellular instability \and rotating detonation wave}

\section{Introduction}
It has been known that extensive research has focused on the intricate dynamics of one/two-dimensional (1D/2D) detonations through numerical investigations. One of the major challenges in multidimensional detonation simulation is the high computational cost associated with accurately resolving the essential detonation structures, as evidenced by 2D studies such as those by \citet{shen2017role}, \citet{han2017role}, and \citet{shi2020re}. While three-dimensional (3D) detonation simulations have also been explored in some details, they remain less comprehensive compared to their 2D counterparts. This discrepancy can be attributed to the significant computational complexity and challenges associated with both performing and analyzing resolved 3D detonation simulations. Early experimental work \citep{takai1975study} and numerical observations \citep{toshi1989propagation,williams1996detailed} shed light on the characteristics of wave structures of 3D detonations. Numerically, \citet{dou2008simulations} documented the presence of spinning motion in narrow ducts and rectangular modes in wider ducts, while \citet{wang2013high} investigated the impact of overdriven factors on highly unstable detonations. Recently, \citet{crane2023three} conducted direct simulations of 3D detonations confined within a square channel or a round tube geometry, incorporating detailed chemical kinetics. The cellular structure observed within the square channel exhibited a higher cell length-width ratio, with the multi-kernels being found much stronger than the line-kernels. \citet{yao2024stereoscopic} examined the 3D detonation in varying square ducts, correlating the 2D and 3D detonation phenomena. Furthermore, \citet{zhang2024numerical} conducted a detailed examination of spinning detonation at exceptionally high resolutions and compared the corresponding wave structures with their 2D counterparts.

While the available studies provide some fundamental insights into the 3D behavior of detonation waves, another critical aspect to consider is their implications in practical applications. In recent years, the rotating detonation engine (RDE) has gained significant attention due to its potential for high combustion efficiency and continuous thrust generation
\citep{rankin2017overview}. A typical RDE topology involves a co-axial combustor, and one or more rotating detonation waves propagate circumferentially inside it. Extensive 2D simulations of RDEs have been conducted \citep{yi2011propulsive,tsuboi2015numerical}, alongside 3D simulations \citep{frolov2013three,smirnov2019three}. Most studies have primarily focused on the macroscopic flow structure and propulsion performance, while the essential understanding of some fundamental detonation physics is still missing. This challenge arises from the multiscale nature of RDE flow dynamics; for instance, while the typical scale of an RDE is around $O\left(10^{-1}\right) \mathrm{m}$, resolving the detonation wave structure requires to consider scales on the order of $O\left(10^{-5}\right) \mathrm{m}$ or even smaller. Consequently, achieving a fully resolved 3D simulation of an RDE demands approximately $O\left(10^{12}\right)$ computational meshes, making it computationally difficult to obtain sufficiently converged results. Nevertheless, a deep understanding of 3D rotating detonation wave dynamics is significant for future modelling and development of high-fidelity RDE.

\begin{figure}[t]
\centering
\includegraphics[width=14cm, trim=0cm 0cm 0cm 0cm, clip]{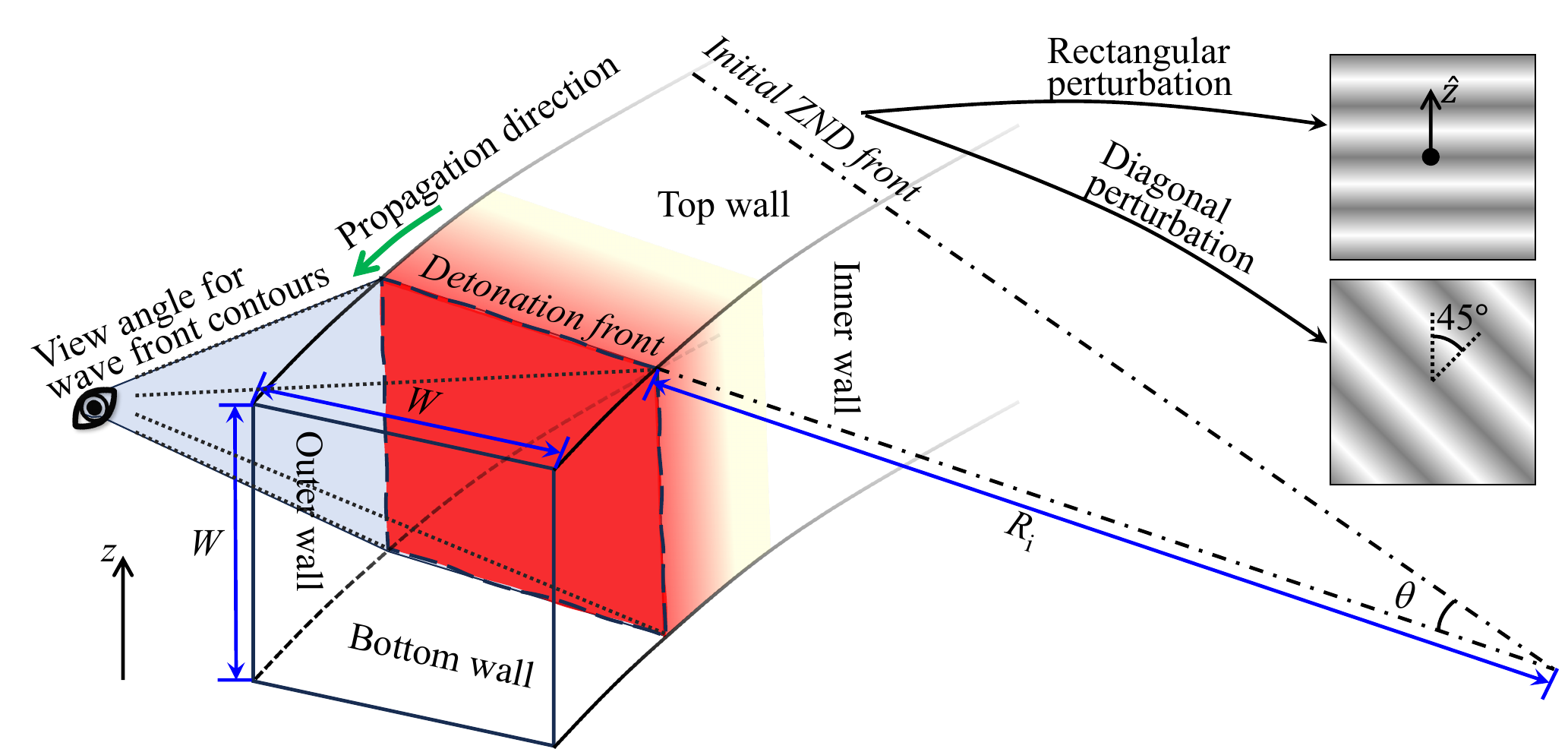}
\caption{Schematic illustrating detonation propagation in a curved channel with a square cross-section. Detonation is deliberately initiated to propagate counterclockwise. The viewing angle in the subsequent figures is positioned in front of the detonation wave front and slightly to the left.}
\label{fig1}
\end{figure}

Additionally, detonation cells appear to correlate with critical conditions necessary to sustain detonation propagation when confined by an inert gas layer \citep{taileb2020influence}. In RDE, the consideration of cell size should account for the 3D effects in curved combustor channels. However, the characteristics of 3D detonation structures within curved ducts remain unresolved. Simplifying the scenario of a 3D RDE by excluding the refilling process and losses due to inert layers, a canonical model arises that a 3D detonation wave propagating along a continuous, elongated curved channel (see Fig. \ref{fig1}, also called rotating detonation wave). Experimental investigations have explored this phenomenon in curved or helical channels \citep{kudo2011oblique,nakayama2012stable,nakayama2013front,pan2018fabrication,rodriguez2019experimental}. Relevant numerical simulations have primarily focused on 2D curved channels. \citet{sugiyama2014numerical} examined detonation waves in a curved 2D channel and proposed critical conditions for stable propagation. \citet{short2019propagation} investigated the effect of weakly unstable detonation wavefronts on cellular structures in a 2D circular arc geometry through high-fidelity simulations. For sufficiently wide arcs, the detonation wavefront near the inner side remains hydrodynamically stable without cellular structures, in some cases accompanied by a stable Mach stem near the outer arc boundary. \citet{yao2023characteristics} conducted a parametric study involving various 2D curved channels with differing inner radii. However, the influence of continuous curved channels on 3D detonation wave structures remains unclear. From both fundamental interest and applied perspective, understanding the characteristics of 3D rotating detonation waves is crucial. This study aims to investigate this issue through high-resolution simulations, offering critical insights into the underlying physics in such scenarios, and elucidating the impact of curved ducts on detonation dynamics.

\section{Physical modelling}
\subsection{Governing equations}
In this study, time-resolved 3D simulations are conducted with a wide range of parameters, necessitating significant computational resources. Following prior research on 3D detonation in straight ducts \citep{dou2008simulations,wang2013high,yao2024stereoscopic}, we employ the non-dimensional one-step reaction model and neglect viscous effects. The 3D reactive Euler equations are given by

$$
\frac{\partial \mathbf{U}}{\partial t}+\frac{\partial \mathbf{F}(\mathbf{U})}{\partial x}+\frac{\partial \mathbf{G}(\mathbf{U})}{\partial y}+\frac{\partial \mathbf{H}(\mathbf{U})}{\partial z}=\mathbf{S}
$$

with

$$
\mathbf{U}=\left[\begin{array}{l}
\rho \\
\rho \mathrm{u} \\
\rho v \\
\rho w \\
E \\
\rho \lambda
\end{array}\right], \mathbf{F}=\left[\begin{array}{l}
\rho u \\
\rho u^{2}+p \\
\rho u v \\
\rho u w \\
(E+p) u \\
\rho \lambda u
\end{array}\right], \mathbf{G}=\left[\begin{array}{l}
\rho v \\
\rho v u \\
\rho v^{2}+p \\
\rho v w \\
(E+p) v \\
\rho \lambda v
\end{array}\right], \mathbf{H}=\left[\begin{array}{l}
\rho w \\
\rho w u \\
\rho w v \\
\rho w^{2}+p \\
(E+p) w \\
\rho \lambda w
\end{array}\right], \mathbf{S}=\left[\begin{array}{l}
0 \\
0 \\
0 \\
0 \\
0 \\
\dot{\omega}
\end{array}\right] \text {. }
$$

Here $\rho, p, u, v, w, \lambda$ and $E$ denote dimensionless density, pressure, the fluid velocities in the $x$, $y$, and $z$ directions, the mass fraction of the reactant and the total energy per unit volume, respectively. The chemical reaction rate follows the Arrhenius equation $\dot{\omega}=-K \rho \lambda e^{-\frac{E_{\mathrm{a}}}{T}}$, where $E_{\mathrm{a}}$ is the activation energy and $K$ is a scaling factor adjusted to ensure that the halfreaction-length $\left(\ell_{1 / 2}\right)$ in the ZND profile is of unit length scale. The ideal gas law in nondimensional form is written as $p=\rho T$, and the total energy per unit volume is given by $E=$ $\frac{p}{\gamma-1}+\frac{1}{2} \rho\left(u^{2}+v^{2}+w^{2}\right)+\lambda \rho Q$. The dimensionless forms of variables with respect to the state of the unburned gas are given by:

$$
\begin{gathered}
\rho=\frac{\rho^{\prime}}{\rho_{0}^{\prime}}, p=\frac{p^{\prime}}{p_{0}^{\prime}}, T=\frac{T^{\prime}}{T_{0}^{\prime}}, u=\frac{u^{\prime}}{\sqrt{R T_{0}^{\prime}}}, v=\frac{v^{\prime}}{\sqrt{R T_{0}^{\prime}}}, w=\frac{w^{\prime}}{\sqrt{R T_{0}^{\prime}}} \\
E_{\mathrm{a}}=\frac{E_{\mathrm{a}}^{\prime}}{R T_{0}^{\prime}}, Q=\frac{Q^{\prime}}{R T_{0}^{\prime}}, x=\frac{x^{\prime}}{L_{1 / 2}^{\prime}} \text { and } t=\frac{t^{\prime}}{L_{1 / 2}^{\prime} / \sqrt{R T_{0}^{\prime}}},
\end{gathered}
$$

where $R$ is the gas constant. The variable with a prime indicates its dimensional form, and the one with a subscript 0 indicates the corresponding value of the unburned gas.

Initially, a quasi-1D detonation profile using the Zel'dovich-von Neumann-Döring (ZND) model is imposed at the far end of the domain, accompanied by perturbations in total energy to trigger cellular instabilities (Fig. \ref{fig1}). Here, we consider two types of perturbations that differ in the direction of perturbation. Specifically, for rectangular perturbations, it is formulated as

$$
E=E \cdot\left(1+p_{\mathrm{amp}} \cdot \cos \left(2 \pi \cdot \frac{\hat{z}}{p_{\mathrm{wav}}}\right)\right)
$$

where $p_{\text {amp }}=0.05$ and $p_{\text {wav }}=2$ represent the amplitude and wavelength of the perturbation, respectively. $\hat{z}$ denotes the distance from the center point of the cross section along the $z$ direction (Fig. \ref{fig1}). Additionally, to initialize cases with diagonal perturbations, the above expression is simply rotated by $45^{\circ}$ in space.

\subsection{Numerical methods}
\label{sec2.2}
The presented study utilizes an in-house parallel computing code to integrate the above equations. The second-order Conservation Element and Solution Element scheme with local Lax-Friedrichs flux as the inner fluxes (LLF-CESE) in a cylindrical coordinate proposed and validated by \citet{shi2023numerical} is used for fitting the curved 3D simulation domain. To further validate the solver for the problems addressed in this study, two well-established cases are presented in the \hyperref[appen]{Appendix}: a 2D detonation in a wide curved channel and a 3D spinning detonation. All simulations are performed with a CFL number of 0.9, and implicit time integration is applied to resolve chemical reactions. The investigation considers ducts with square cross-sections of side lengths ($W$) of 10 or 20 length-units. Straight ducts and curved ducts with inner radii of 200, 100, and 50 length-units are analyzed. These setups aim to investigate the effects of geometric configurations on detonation dynamics under varying levels of curvature confinement.

\begin{figure}[t]
\centering
\includegraphics[width=13cm, trim=0cm 0cm 0cm 0cm, clip]{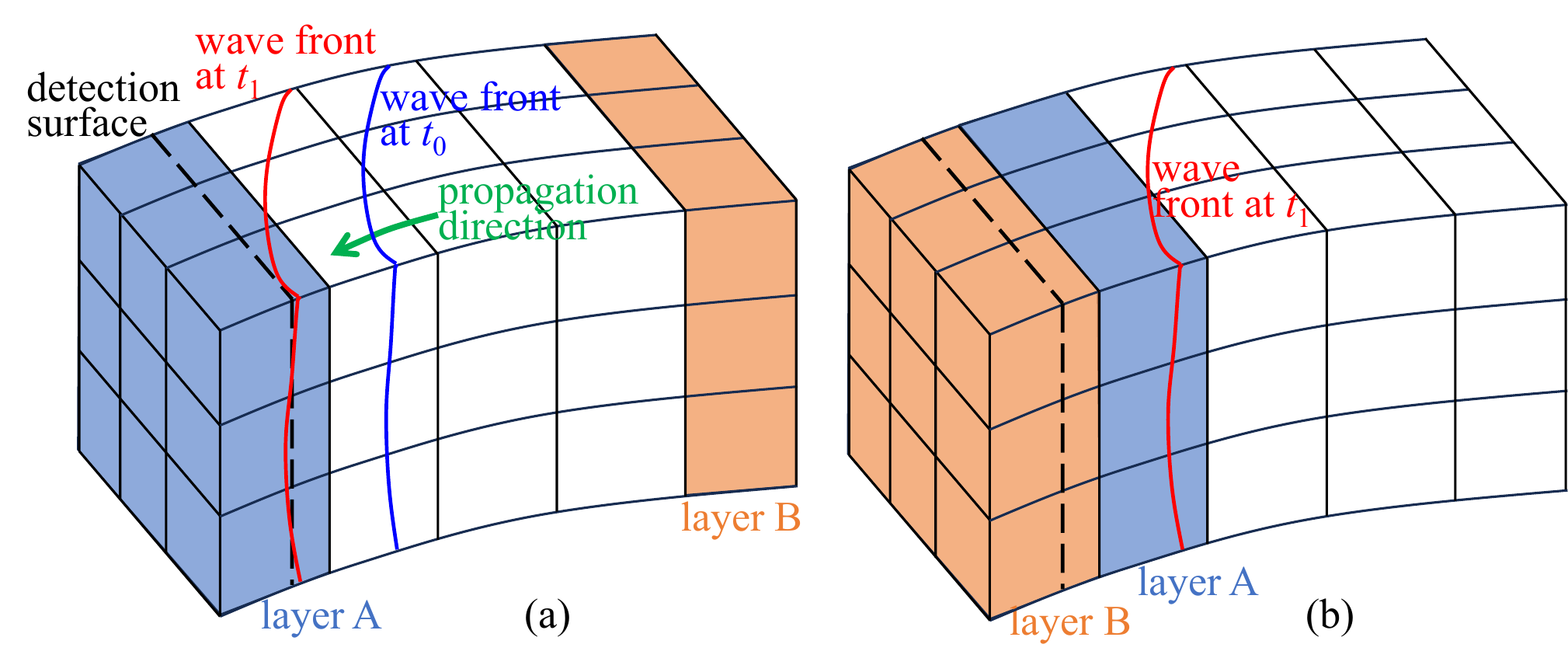}
\caption{Schematic of moving domain technique with domain decomposition in a 3D curved duct. Each small cubic represents a subordinate computational domain. The dashed line represents the detection surface of shock arrival when a domain re-arrangement operation is needed. (a) Before domain re-arrangement, subordinate domain layer $A$ and $B$ are at the two ends of the domain, the wave front reach the detection surface at $t_1$; (b) After domain re-arrangement, layer $B$ is now reinitiated and attached in front of layer $A$.}
\label{fig2}
\end{figure}

Furthermore, 3D detonation waves usually require a long propagation distance to reach a fully developed state. In previous studies of 3D detonation in straight tubes \citep{dou2008simulations,wang2013high,crane2023three}, inflow conditions are commonly set to ChapmanJouguet (CJ) or overdriven velocities, with corresponding free outflow boundaries, which effectively constrain the detonation wave within the domain. However, applying these methods to curved channels presents challenges. Firstly, the movement of the channel wall relative to the detonation wave's frame is non-parallel (i.e., the wave's global propagation direction is non-parallel to the channel wall), which complicates the implementation of proper inflow boundary conditions. Secondly, a notable velocity deficit, defined as the ratio of local normal detonation velocity to the theoretical CJ value ($D / D_{\mathrm{CJ}}$), exists particularly near the inner wall, and there is no definitive approach to predetermine this velocity. Therefore, in this study, a moving domain technique is adopted instead where the simulation domain tracks and encompasses the detonation waves and requisite wake flow consistently \citep{sow2019viscous,nejaamtheen2021effects}. Meanwhile, this approach is integrated with domain decomposition techniques (depicted in Fig. \ref{fig2}). For parallelization with Message Passing Interface (MPI), the moving domain technique is accomplished by assigning proper neighboring information for each subordinate domain and exchange overlapped data in each subordinate. When the detonation wave advances beyond a certain point in the front layer of subordinate domains, the back layer is numerically reattached to the front layer and reset with initial unreacted gas conditions as the new front layer. In this way, the moving domain is efficiently implemented without copying massive flow data among subordinate domains, and is ensured to progress at the detonation velocity (not necessarily CJ speed). The moving domain can accommodate various conditions without wave reflection at the far end boundary, and extend the computational domain to arbitrary lengths (not limited to less than one circular cycle).

Adiabatic slip boundary conditions are applied to the top, bottom, inner and outer walls. All the cases are examined to ensure that the result is not affected by the computational setup. Notably, the cases with a straight duct can be regarded as a curved duct with an infinite inner radius ($R_{\mathrm{i}}=\infty$). In practice, an inner radius of $R_{\mathrm{i}}=10^{10}$ is utilized to approximate infinite inner radius, effectively representing a straight duct (for convenience, the terms ``inner wall'' and ``outer wall'' are still used to refer to the two side walls in straight duct scenarios). In the following text, the terms ``straight duct'' and ``duct with $R_{\mathrm{i}}=\infty$'' are used interchangeably to represent the same concept. These computations are performed in the Tianhe supercomputer in Tianjin Supercomputer Center, China.

\section{Numerical results}
In the present study, all simulations are analyzed when the wave structures are fully developed. These numerical configurations are summarized in Tab. \ref{tab1}, encompassing variations in inner radii ($R_{\mathrm{i}}$), initial perturbations, mesh resolutions ($\Delta h$), side lengths ($W$) of duct cross-section, and activation energies ($E_{\mathrm{a}}$). At the mesh resolution of $\Delta h=1 / 20$, the number of computational cells clustered to near the detonation wave front for the narrow duct case ($W=10$) is $200 \times 200 \times 400$, and that for the wide duct case ($W=20$) is $400 \times$ $400 \times 800$. The detonation is consistently initiated to propagate counterclockwise. Therefore, in all subsequent figures where the detonation wave front faces the reader, the left side corresponds to the outer wall, while the right side pertains to the inner wall (Fig. \ref{fig1}). Without explicitly stating it, the bulk detonation wave velocity is defined as the average velocity measured along the inner wall over a substantial distance. Throughout the simulations, numerical soot foils are recorded based on the maximum pressure history. To ensure consistency across presentations, all soot foils display maximum pressures ranging from 10 to 100 in grayscale. In the following section, we first focus on examining the impacts of curvatures and initial perturbations on 3D detonations within narrow ducts. Subsequently, we progress to investigate scenarios with a wider duct, and explore the effects of higher activation energies particularly.

\begin{table}[t]
\centering
\caption{Setups of numerical simulations in the present study.}
\begin{tabular}{@{\hskip 0pt}lcccccc@{\hskip 0pt}}
\hline
\textbf{Case} & $E_{\mathrm{a}}$ & $R_{\mathrm{i}}$ & $W$ & $\Delta h$ &Perturbation&Studied parameter \\ 
\hline
Case $1^a$ & 10 & $\infty$ & 10 & 1/20 & Rectangular & \multirow{4}{*}{\begin{tabular}{c}Inner radius\end{tabular}} \\
Case $2^a$ & 10 & 200 & 10 & 1/20 & Rectangular & \\
Case $3^a$ & 10 & 100 & 10 & 1/20 & Rectangular & \\
Case $4^a$ & 10 & 50 & 10 & 1/20 & Rectangular & \\
\hline
Case 5 & 10 & $\infty$ & 10 & 1/20 & Diagonal & \multirow{4}{*}{\begin{tabular}{c}Initial perturbation\end{tabular}} \\
Case 6 & 10 & 200 & 10 & 1/20 & Diagonal & \\
Case 7 & 10 & 100 & 10 & 1/20 & Diagonal & \\
Case 8 & 10 & 50 & 10 & 1/20 & Diagonal & \\
\hline
Case 9 & 10 & 50 & 10 & 1/10 & Rectangular & \multirow{2}{*}{\begin{tabular}{c}Mesh resolution\end{tabular}} \\
Case 10 & 10 & 50 & 10 & 1/30 & Rectangular & \\
\hline
Case 11 & 10 & $\infty$ & 20 & 1/20 & Rectangular & \multirow{4}{*}{\begin{tabular}{c}Cross-\\section size\end{tabular}} \\
Case $12^{b}$ & 10 & 200 & 20 & 1/10 & Rectangular & \\
Case $13^{b}$ & 10 & 100 & 20 & 1/10 & Rectangular & \\
Case 14 & 10 & 50 & 20 & 1/20 & Rectangular & \\
\hline
Case $15^{b}$ & 20 & $\infty$ & 20 & 1/10 & Rectangular & \multirow{4}{*}{\begin{tabular}{c}Activation energy\end{tabular}} \\
Case $16^{b}$ & 20 & 50 & 20 & 1/10 & Rectangular & \\
Case 17 & 50 & $\infty$ & 20 & 1/20 & Rectangular & \\
Case 18 & 50 & 50 & 20 & 1/20 & Rectangular & \\
\hline
\end{tabular}\\
\noindent\textsuperscript{$a$} \small{Under similar conditions, a 2D simulation is also performed.}\\
\noindent\textsuperscript{$b$} \small{A moderate mesh resolution is used for obtaining the detonation velocity only.}\\
\label{tab1}
\end{table}

\subsection{Effects of curvature in narrow ducts (cases 1-4)}
\label{sec3.1}
\begin{figure}[t]
\centering
\includegraphics[width=13cm, trim=0cm 0cm 0cm 0cm, clip]{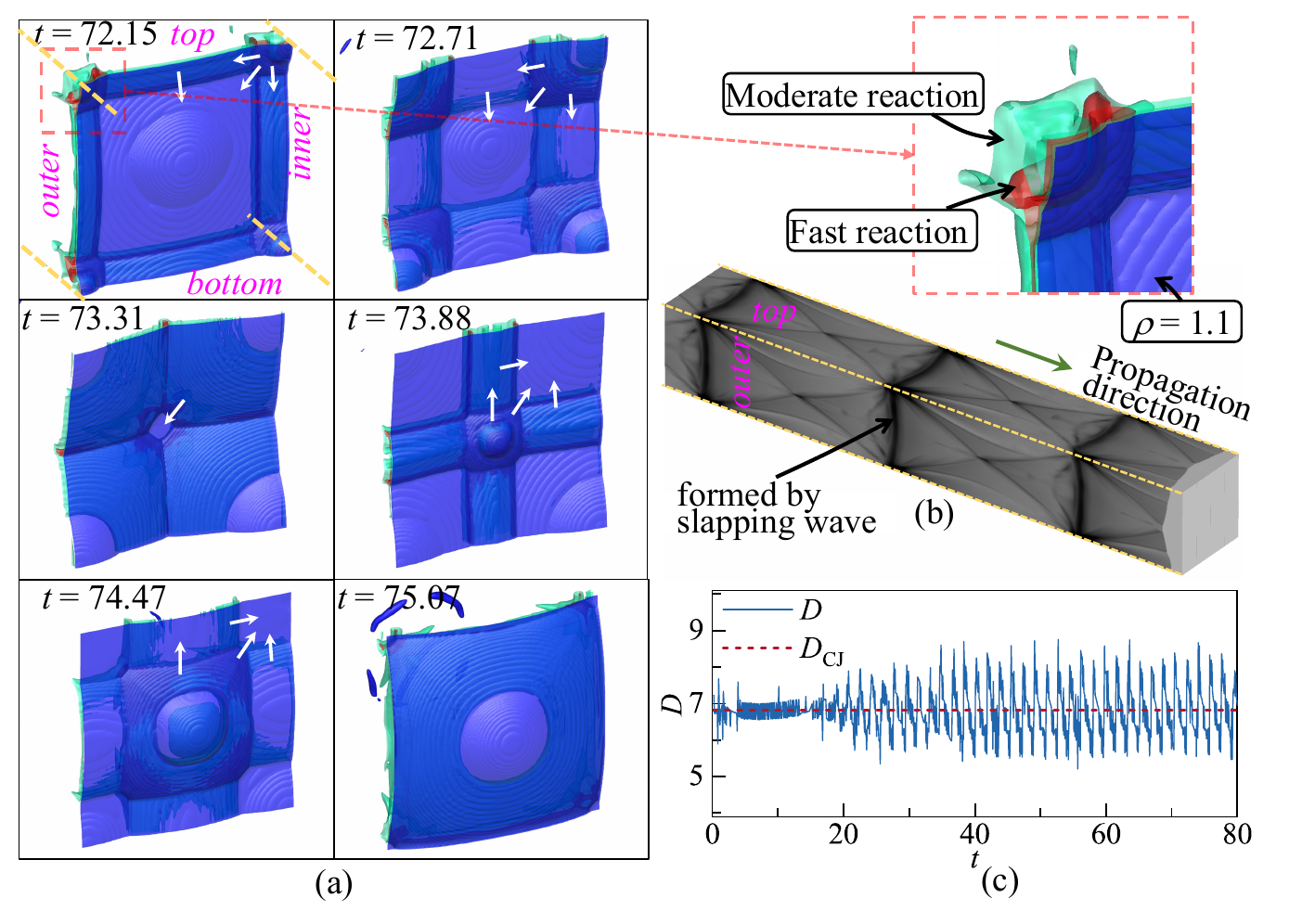}
\caption{Case 1: (a) 3D views of wave fronts and chemical reactions during detonation propagation in a straight duct (refer to the view angle in Fig. \ref{fig1}). The dashed yellow lines indicate the edges of the duct. The blue iso-surface, defined by $\rho$ = 1.1, represents the detonation front, while the red and green iso-surfaces denote the fast and moderate chemical reaction zones, providing a rough indication of blast kernels. Representative wave directions are marked with white arrows for clarity. (b) Numerical soot foil. (c) Evolution of bulk velocity.}
\label{fig3}
\end{figure}

In this section, all cases are initialized with rectangular perturbations. After propagating over a significant distance to fully develop the detonation structures, the straight duct case ($R_{\mathrm{i}}=\infty$) exhibits a single internal cell with two transverse waves on each wall. The cellular structures formed on adjacent side walls are phase-synchronized, causing transverse waves from neighbouring walls to simultaneously collapse near the corner lines. Figure \ref{fig3}a illustrates these wave structures within one propagation cycle. At $t=72.15 \sim 72.71$, symmetric pairs of transverse waves propagate towards each other, forming four bulges near the corners due to their interactions. Fast reactions occur behind the transverse waves and Mach stems, defined as ignition kernels per \citet{crane2023three}, while moderate reactions follow the incident shocks. During $t=73.31 \sim 74.47$, these transverse waves collide near the cross-section center, reflect, and then propagate back towards the walls. By $t=75.07$, these waves reach the walls, generating high-pressure regions as depicted in the numerical soot foil (Figure \ref{fig3}b), where dark areas indicate elevated pressure. The soot foil reveals trajectories of triple lines formed by interactions of incident shocks and transverse waves and highlights the significant impact of the slapping waves. Furthermore, compression waves originating from corners where transverse waves collide are evident. These transverse waves subsequently reflect to reproduce similar wave structures observed earlier at $t=72.15$, completing a full propagation cycle. The bulk velocity within the rectangular duct closely matches theoretical values, with instantaneous fluctuations of detonation velocity due to the small duct size and periodic oscillations of transverse waves (Figure \ref{fig3}c).

\begin{figure}[t]
\centering
\includegraphics[width=11.5cm, trim=0cm 0cm 0cm 0cm, clip]{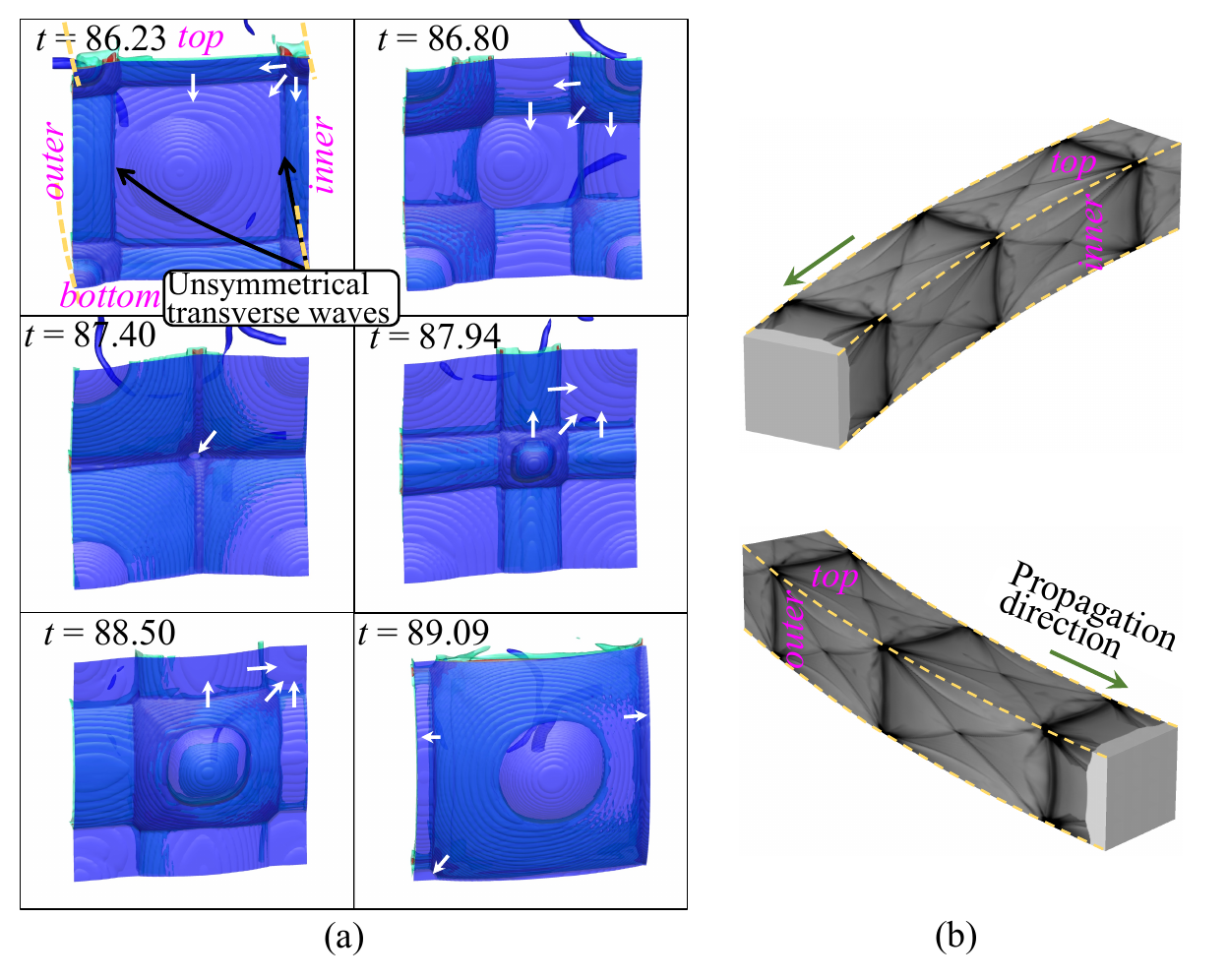}
\caption{Case 2: (a) Wave structures during propagation in a curved duct with $R_{\mathrm{i}}=200$. (b) Soot foil from two viewing angles.}
\label{fig4}
\end{figure}

For curved duct with $R_{\mathrm{i}}=200$, the transverse waves in the horizontal directions do not exhibit symmetry relative to the central plane (Fig. \ref{fig4}a), despite overall wave motions closely resembling those observed in the straight duct case. As a result, in Fig. \ref{fig4}b, the top and bottom walls exhibit a slightly unsymmetrical cellular structure, yet the waves on neighboring walls remain in phase. The bulk velocity measured along the inner wall is only slightly smaller than the CJ value, at around $98 \% D_{\text {CJ }}$.

\begin{figure}[t]
\centering
\includegraphics[width=15cm, trim=0cm 0cm 0cm 0cm, clip]{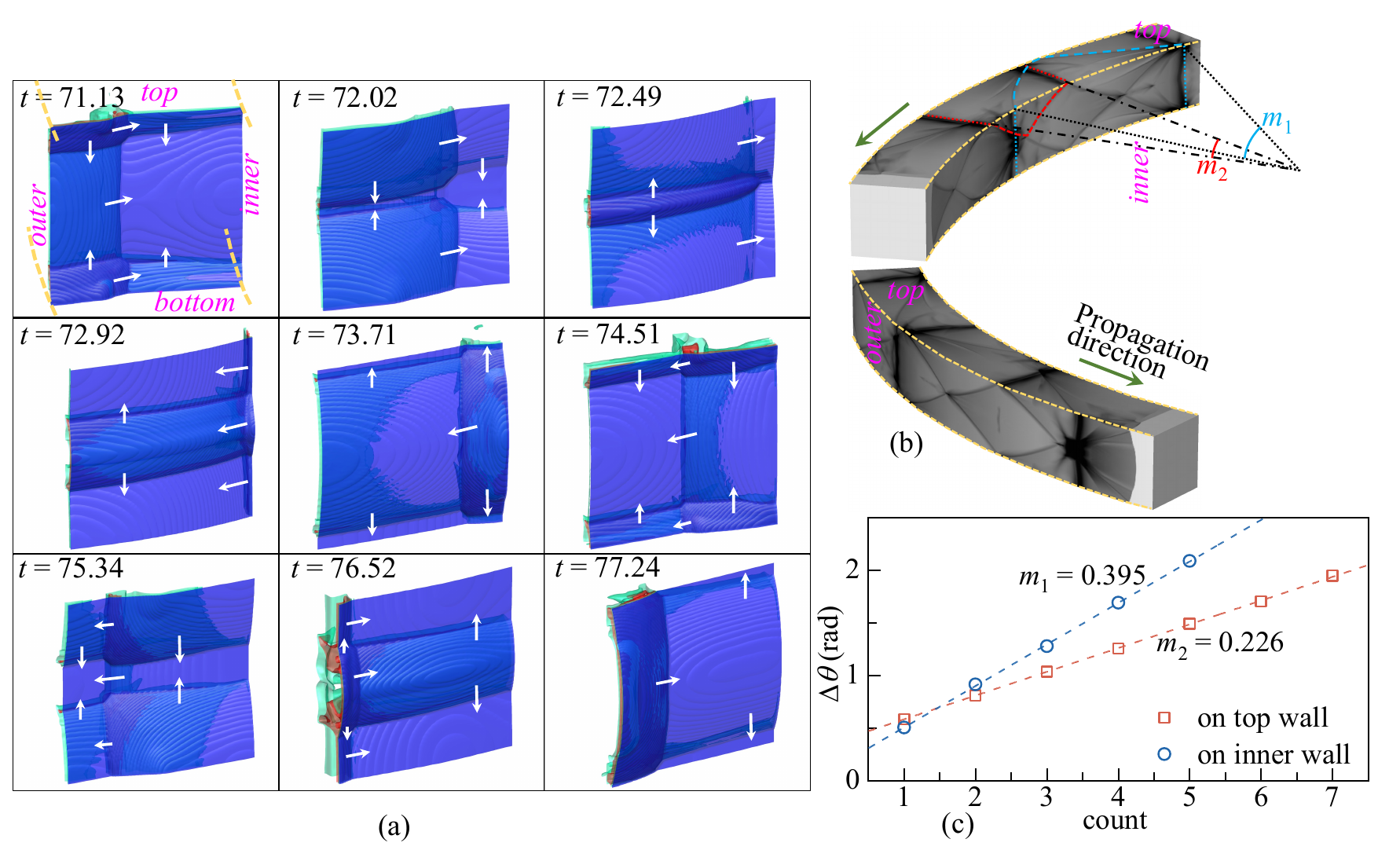}
\caption{Case 3: (a) Wave structures during propagation in a curved duct with $R_{\mathrm{i}}=100$. (b) Soot foil from two viewing angles, with red and blue curves representing the trajectories of transverse waves and slapping waves. (c) Distributions of slapping waves on the top and side walls using the referenced angle $(\Delta \theta)$. The slopes $(m)$ of the fitted lines indicate the angles between two successive slapping wave imprints.}
\label{fig5}
\end{figure}

For curved duct with $R_{\mathrm{i}}=100$, the wave structure exhibits significant differences compared to both the straight duct and $R_{\mathrm{i}}=200$ cases. There is a pair of transverse waves in the vertical direction (parallel to the inner and outer walls), whereas only one transverse wave is present in the horizontal direction (parallel to the upper and bottom walls) (Fig. \ref{fig5}a). Additionally, it is noteworthy that the periods of the horizontal and vertical waves differ (Fig. \ref{fig5}b), the slapping waves recorded by soot foil on the top and bottom walls do not synchronize with those on the side walls. The measured distributions of slapping waves on top and inner walls in terms of angle from an arbitrary reference point show that the occurrence of slapping waves on these two sides are essentially independent of each other (Fig. \ref{fig5}c).

\begin{figure}[t]
\centering
\includegraphics[width=15cm, trim=0cm 0cm 0cm 0cm, clip]{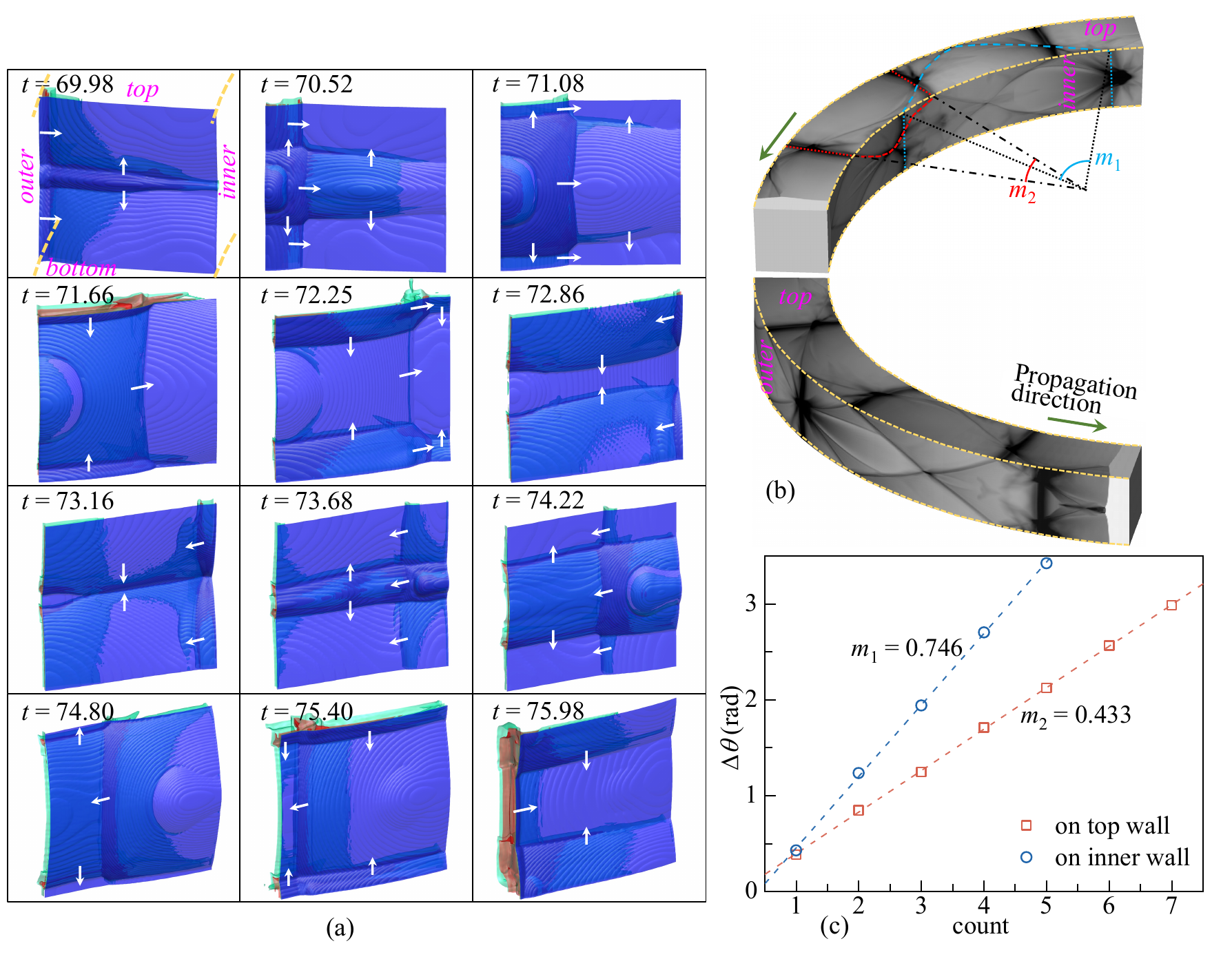}
\caption{Case 4: (a) Wave structures during propagation in a curved duct with $R_{\mathrm{i}}=50$. (b) Soot foil from two viewing angles. (c) Distributions of slapping waves on the top and side walls.}
\label{fig6}
\end{figure}

\begin{figure}[t]
\centering
\includegraphics[width=14cm, trim=0cm 7cm 0cm 0cm, clip]{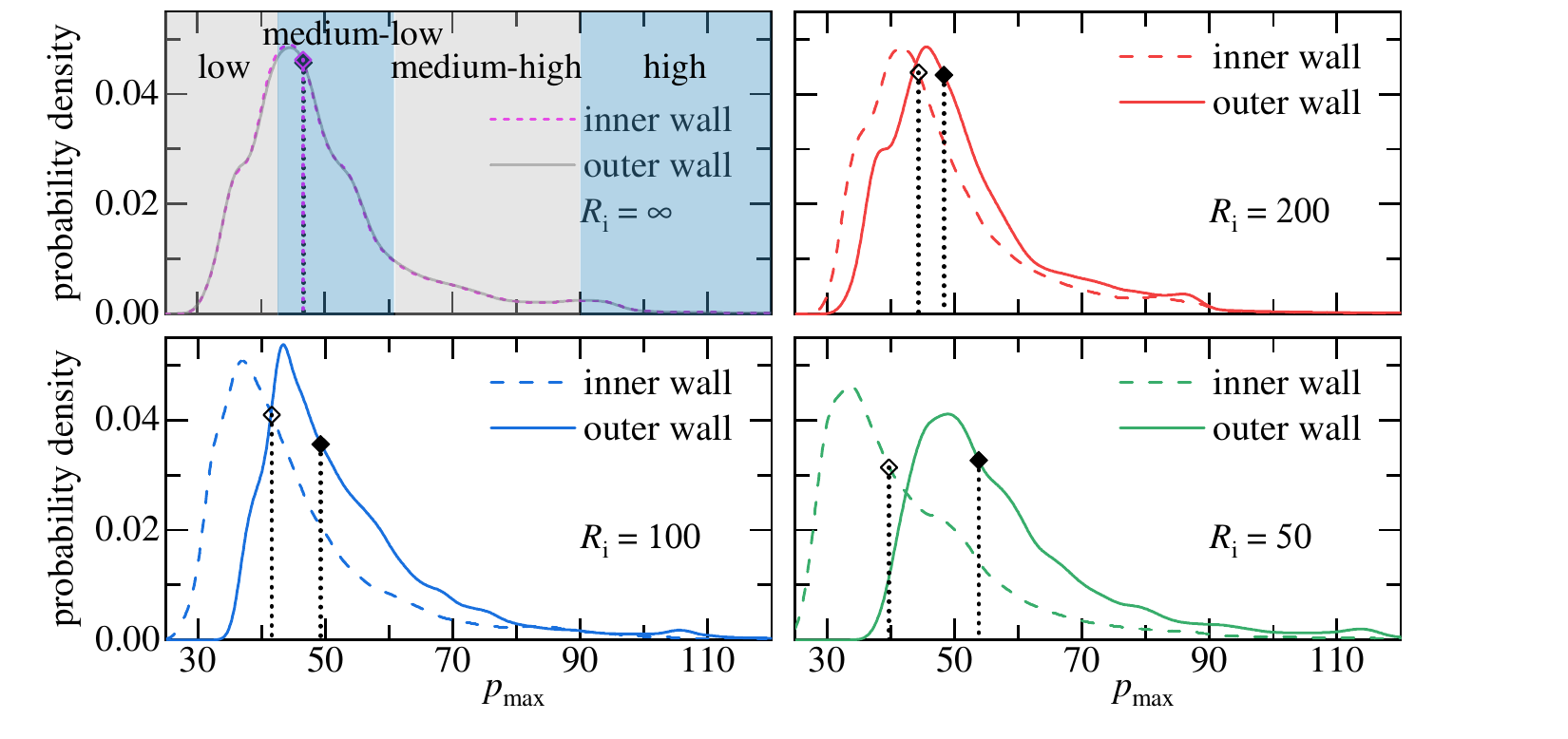}
\caption{Histograms of the maximum pressure recorded on the inner and outer walls for various inner radii. The symbols with drop lines indicate the median values.}
\label{fig7}
\end{figure}

For curved duct with $R_{\mathrm{i}}=50$, the macroscopic wave structures (Fig. \ref{fig6}) are in similar pattern to the case with that with $R_{\mathrm{i}}=100$. To quantify the variations across different inner radii, the histograms depicting the maximum pressure ($p_{\max }$) on the side walls are presented in Fig. \ref{fig7}. For the straight duct, these histograms are nearly indistinguishable on both walls since the wave structures are perfectly symmetrical. The distribution in the low region is primarily influenced by incident shocks, while the medium-low region results from transverse waves (characterized by triple point trajectories on the wall). The medium-high region is affected by slapping waves, and the upper high-pressure range is shaped by shocks converging near the duct corners. In the case of curved ducts, the histogram illustrating $p_{\max }$ on the outer wall exerts fewer fractions of low values. Specifically, the median $p_{\max }$ on the outer wall increases with decreasing inner radius $R_{\mathrm{i}}$. Conversely, the histogram for $p_{\max }$ on the inner side wall shifts left compared to that of the outer side wall, a difference accentuated at smaller $R_{\mathrm{i}}$. The variation in $p_{\max }$ primarily stems from the compression of the detonation wave against the outer side wall during propagation, leading to elevated pressures, particularly noticeable at smaller $R_{\mathrm{i}}$. In contrast, detonation locally diffracts near the inner side wall and thus leading to lower $p_{\text {max }}$.

\subsection{Effects of initial perturbation (cases 5-8)}
\begin{figure}[t]
\centering
\includegraphics[width=13cm, trim=0cm 0cm 0cm 0cm, clip]{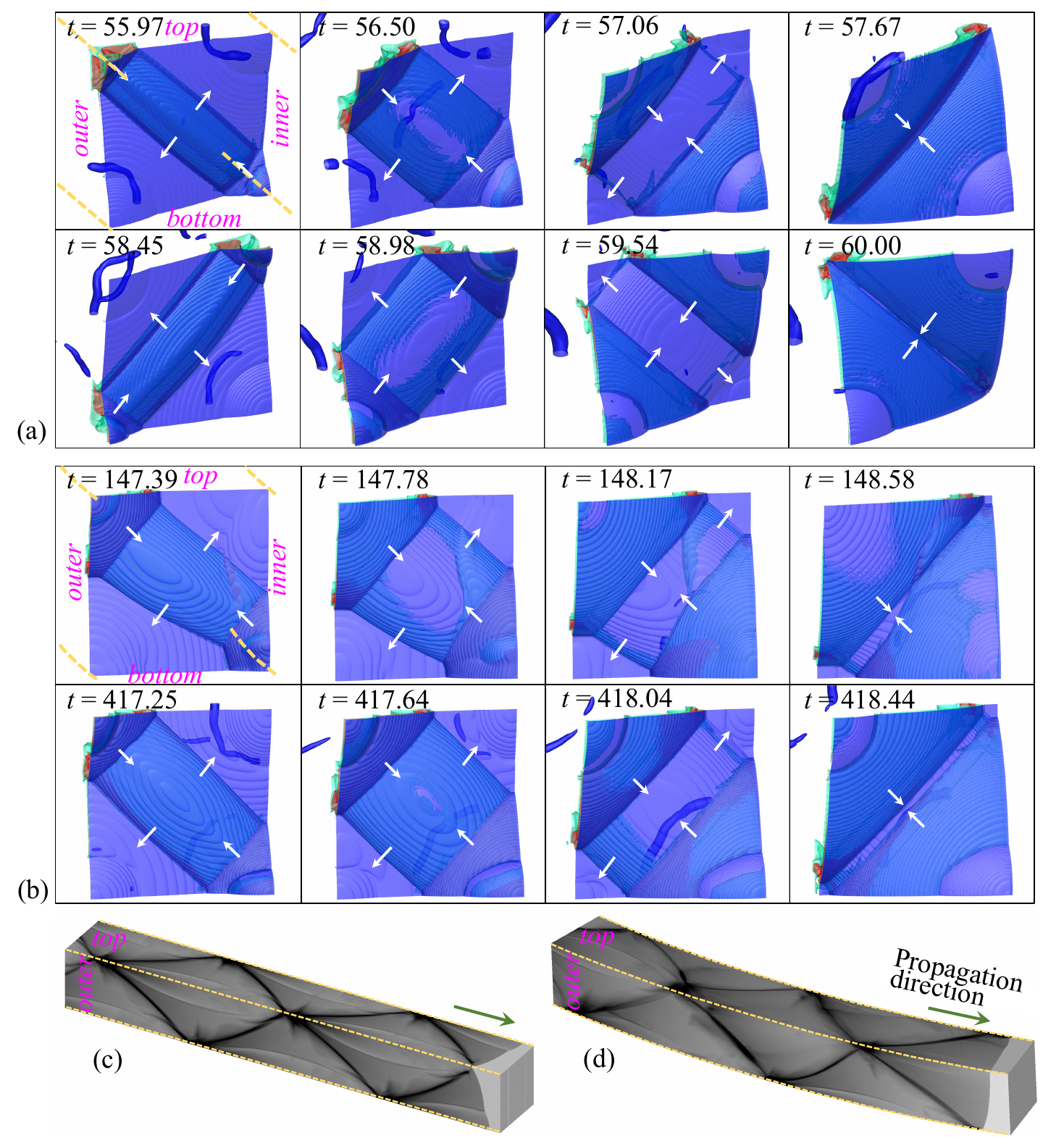}
\caption{Wave structures during propagation in a duct: (a) Case 5: $R_{\mathrm{i}} = \infty$ and (b) Case 6: $R_{\mathrm{i}} = 200$. Soot foils for (c) Case 5: $R_{\mathrm{i}} = \infty$ and (d) Case 6: $R_{\mathrm{i}} = 200$.}
\label{fig8}
\end{figure}

It is established that within a straight duct, detonation can propagate in two distinct modes: transverse waves travel either along the horizontal or vertical direction (rectangular mode, as demonstrated in Sec. \ref{sec3.1}), or diagonally (diagonal mode) \citep{hanana2001pressure,zhang2001direct,he2001three,ji2022three}. Here we investigate the behaviour of detonation waves in response to alternative initial perturbations. Specifically, we initiate diagonal perturbations in the aforementioned scenarios and analyse their respective responses. As shown in Fig. \ref{fig8}a and \ref{fig8}c, in a straight duct where transverse waves propagate diagonally, there are no observable slapping waves on the side walls since the collisions of transverse waves occur along the diagonal lines inside the duct. It is noteworthy that these two sets of transverse waves propagate in opposite phases: as one pair advance towards the center of the cross section, the other pair moves towards the corners.

For the case with $R_{\mathrm{i}}=200$, the early evolution resembles the $R_{\mathrm{i}}=\infty$ case but is slightly influenced by the curvature of the duct, eventually transitioning into the diagonal mode. However, achieving full development of the diagonal mode requires a significantly greater distance. This progression is illustrated in Fig. \ref{fig8}b and \ref{fig8}d, where a global diagonal mode is visible around $t \sim 148$. By $t \sim 420$, this mode becomes much clearer and more structured as the detonation wave completes more than two circular rounds.

\begin{figure}[t]
\centering
\includegraphics[width=13cm, trim=0cm 0cm 0cm 0cm, clip]{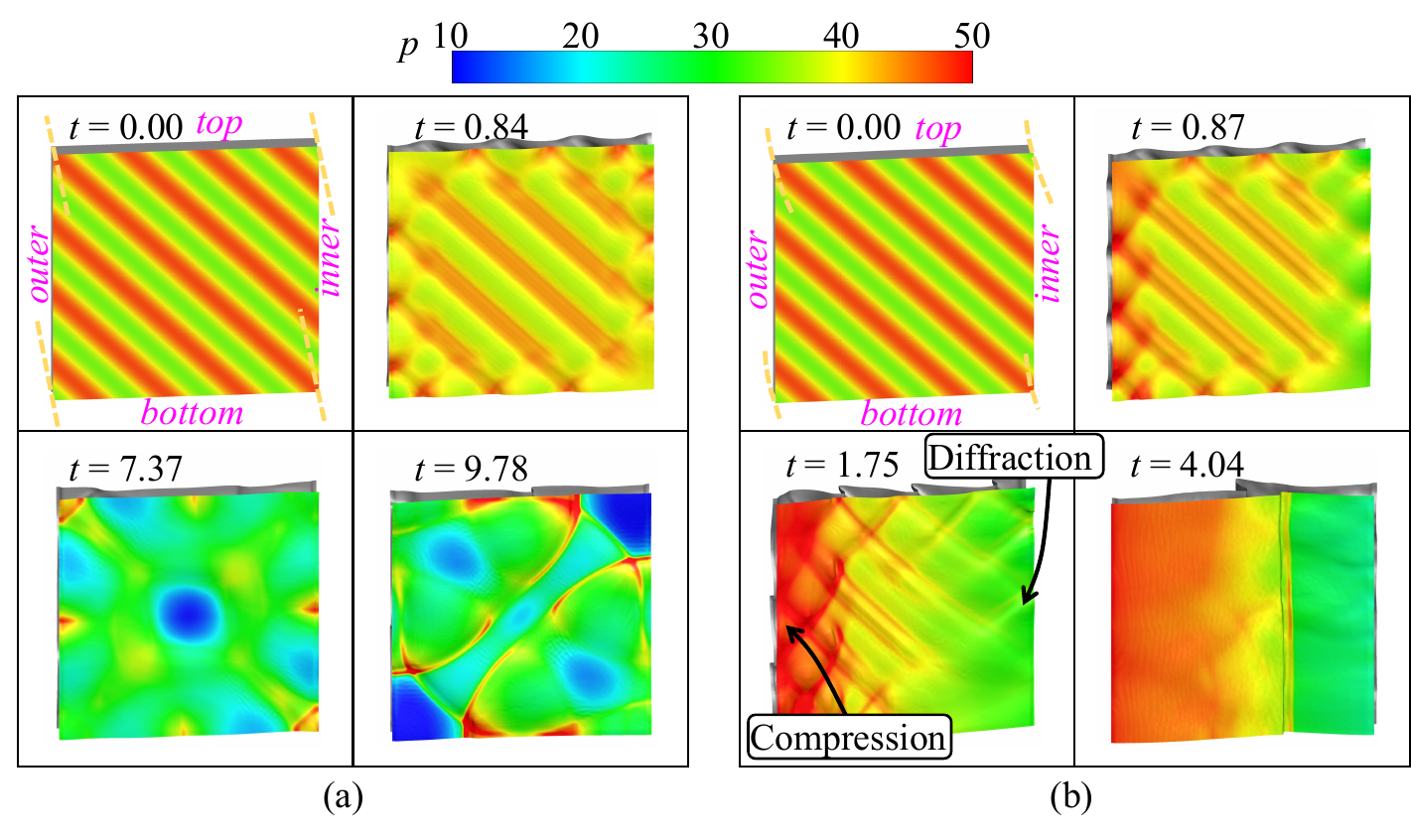}
\caption{Early evolution of diagonal perturbation with (a) Case 5: $R_{\mathrm{i}} = \infty$ and (b) Case 8: $R_{\mathrm{i}} = 50$. The first layer shows the iso-surface of $\lambda = 0.9$ with pressure contours, while the second gray layer behind represents the iso-surface of $\lambda = 0.5$.}
\label{fig9}
\end{figure}

For scenarios with diagonal perturbations and $R_{\mathrm{i}}=100$ or $R_{\mathrm{i}}=50$, however, the detonation waves quickly transition into the rectangular mode. The early-stage evolutions for $R_{\mathrm{i}}=\infty$ and $R_{\mathrm{i}}=50$ are illustrated in Fig. \ref{fig9}. Initially at $t=0$, the detonation wave is perturbed along the diagonal direction as detailed in Sec. \ref{sec2.2}. In the case with $R_{\mathrm{i}}=\infty$, the evolution of wave structures exhibits nearly symmetric behavior relative to the diagonal lines (from $t=0.84$ to 9.78), where these minor perturbations eventually develop into large transverse wave motion along diagonal directions. However, for the case with $R_{\mathrm{i}}=50$, even in the early stages, due to the reflection off the outer wall and expansion near the inner wall, a strong compression wave forms and propagates horizontally toward the inner wall (from $t=$ 0.87 to 1.75). This compression wave quickly transforms into a strong transverse wave, evident from the compact reaction zone trailing behind it (at $t=4.04$). Consequently, the initial perturbation is rapidly influenced by the strong compression and diffraction induced by the curvature of the duct. Following this transition, the propagation speed and wave structures closely resemble those generated by rectangular perturbations, discussed previously in Fig. \ref{fig5} and Fig. \ref{fig6}, and hence are omitted here for brevity.

\begin{figure}[t]
\centering
\includegraphics[width=7cm, trim=0cm 0cm 0cm 0cm, clip]{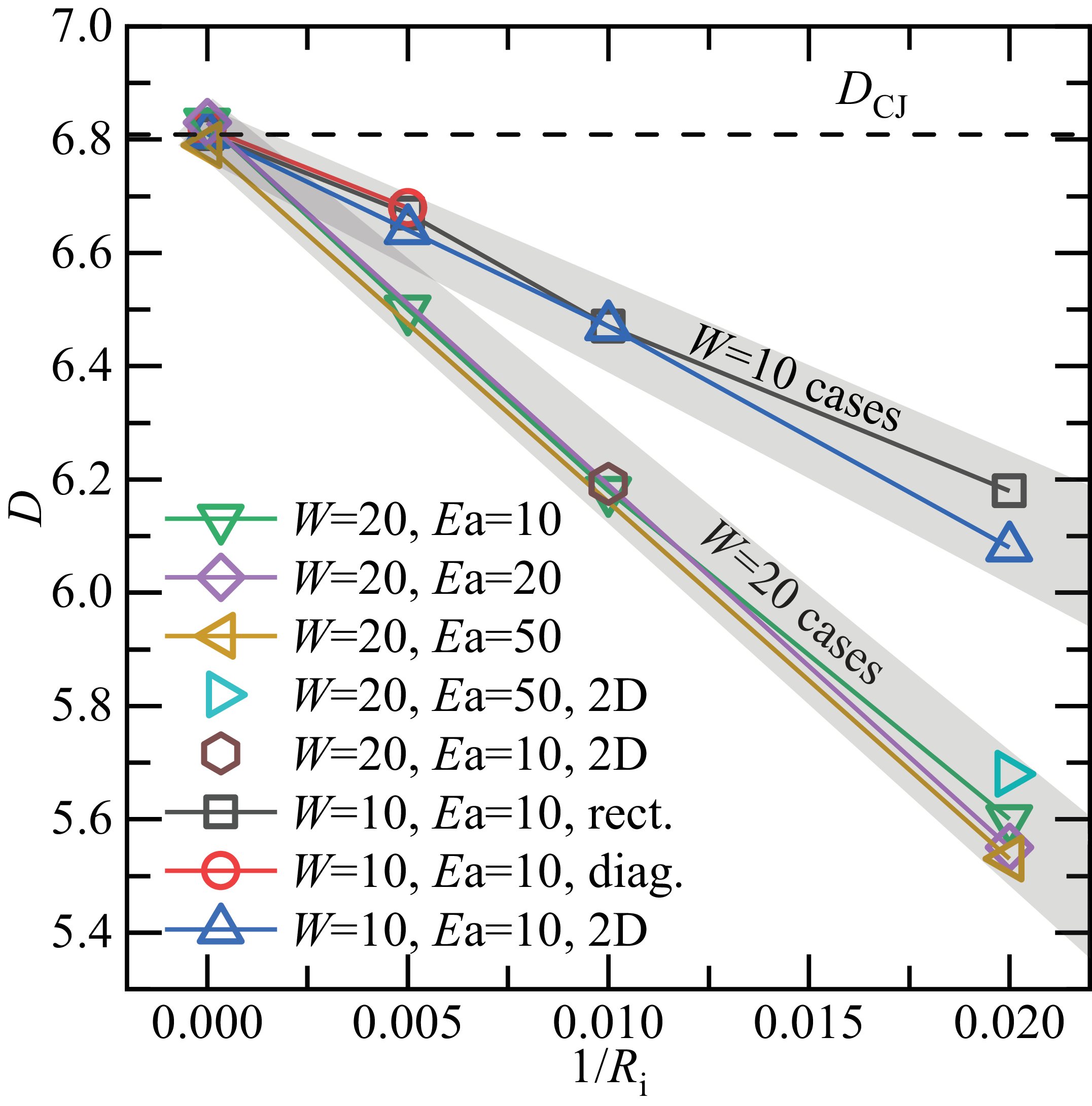}
\caption{Averaged detonation velocity measured along the inner wall.}
\label{fig10}
\end{figure}

In summary, for a diagonal perturbation in a straight duct, there is a rapid transition into the diagonal mode. With a moderate $R_{\mathrm{i}}$, the diagonal mode can exist but requires a longer time to fully develop. For a sufficient small $R_{\mathrm{i}}$, the compression wave and local diffraction due to the curved wall supersede the initial perturbations, leading quickly to the rectangular mode. Nonetheless, the type of propagation mode does not influence the propagation velocities in either straight or curved ducts (as shown by the gray squares and red circles in Fig. \ref{fig10}). It's worth mentioning that while diagonal mode can be initiated by diagonal perturbations in some cases, it can also occur under rectangular perturbations, as will be particularly demonstrated later in Sec. \ref{sec3.5}. This highlights the complex interplay between initial conditions and the eventual mode of detonation wave propagation.

\subsection{Comparison between 2D and 3D simulations (cases 1$^{a}$-4$^{a}$ and 1-8)}
\begin{figure}[t]
\centering
\includegraphics[width=13cm, trim=0cm 2cm 0cm 0cm, clip]{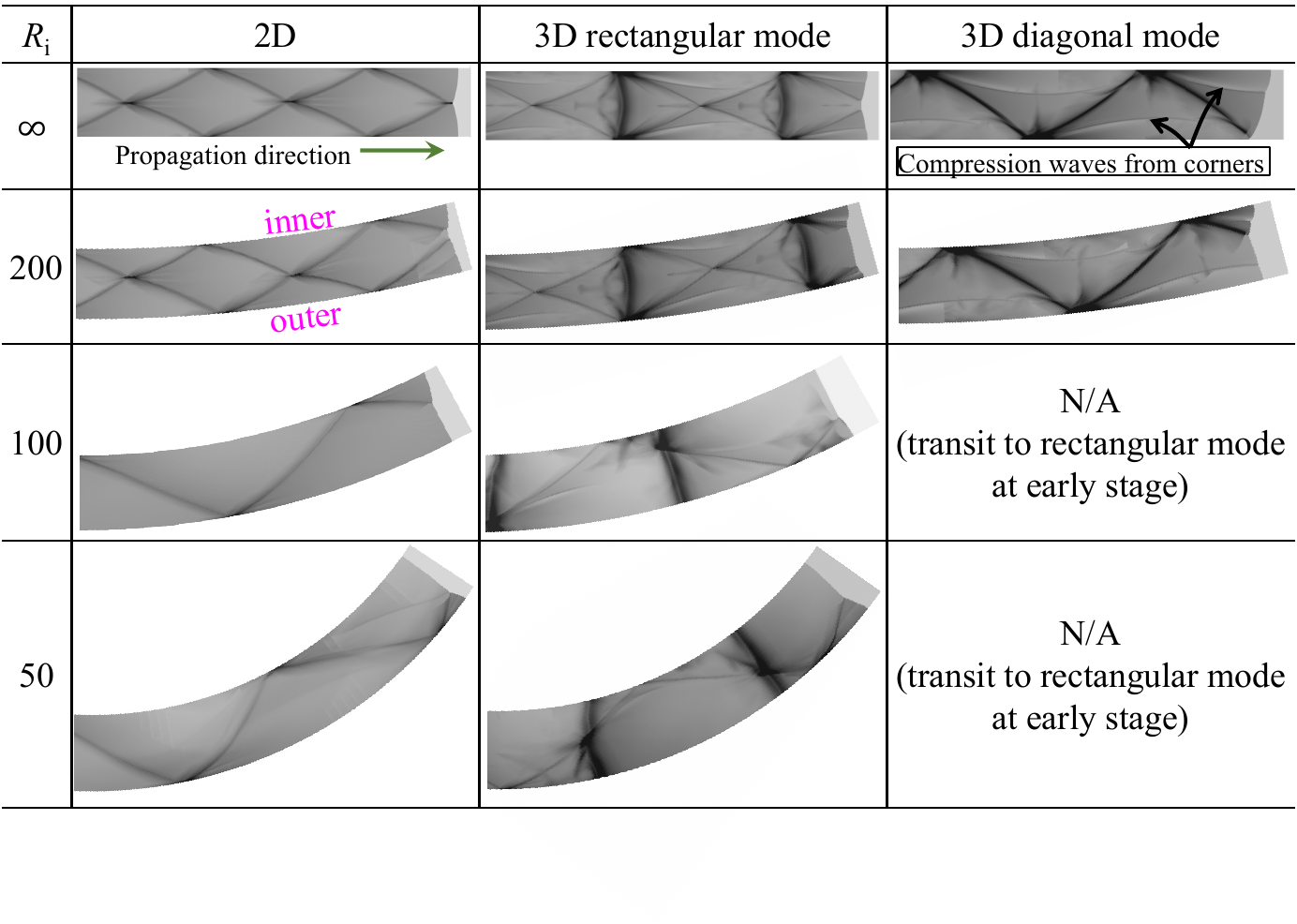}
\caption{Comparison of soot foils from 2D simulations (Cases $1^a$-$4^a$) and soot foils from the top wall of 3D simulations (Cases 1-8) for various $R_{\mathrm{i}}$.}
\label{fig11}
\end{figure}

Here, several 2D simulations are performed with similar conditions except the $z$ dimension diminishes. A comparison between 2D and 3D simulations reveals several significant distinctions (see Fig. \ref{fig11}). First, in the case of a straight duct, the 3D rectangular mode demonstrates cellular structures that closely resemble those observed in the 2D simulations, with comparable cell dimensions. However, a prominent disparity arises in the form of a dark region caused by the slapping wave, which has also been distinctly observed in experiments \citep{monnier2022analysis}. A less conspicuous secondary difference lies in the compression wave emanating from the 3D corners. Conversely, in the case of a straight duct employing the diagonal mode, there is an absence of a slapping wave, as all collisions of transverse waves occur exclusively at the diagonal surfaces. Nevertheless, the presence of a compression wave originating from the corners remains observable. Additionally, only one transverse wave is noted on the upper wall in diagonal mode. The situation remains largely consistent at $R_{\mathrm{i}}=200$, albeit slightly skewed compared to straight configurations. Upon further reduction to $R_{\mathrm{i}}=100$, the compression waves near the outer corner are significantly stronger owing to the larger angle relative to the wall. This angle amplifies further at $R_{\mathrm{i}}=50$, almost blending the compression wave with the transverse wave. Nonetheless, velocities measured in both 2D and 3D simulations with identical $R_{\mathrm{i}}$ values exhibit comparable values, as depicted in Fig. \ref{fig10}.

\subsection{Effects of mesh resolution (cases 9\&10)}
\begin{figure}[t]
\centering
\includegraphics[width=10cm, trim=0cm 0cm 0cm 0cm, clip]{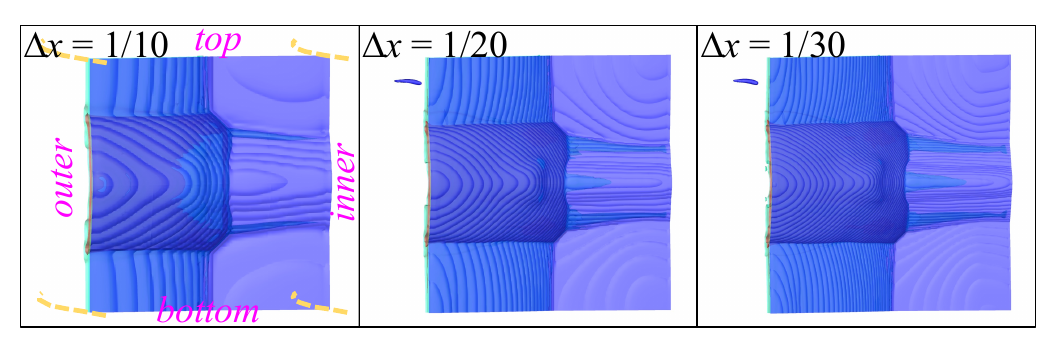}
\caption{Wave structures at the same phase for different mesh resolutions (Case 9, 4, and 10) in the case of $R_{\mathrm{i}} = 50$.}
\label{fig12}
\end{figure}

The resolution of $20 \mathrm{pts} / l_{1 / 2}$ is widely acknowledged as providing high resolution suitable even for 2D simulations, as highlighted in studies by \citet{shen2017role}, \citet{short2019propagation}, and \citet{shi2020re}. Here, we demonstrate the impact of mesh resolution on cases involving highly curved ducts. Figure \ref{fig12} illustrates the wave structures observed across three different mesh resolutions. All configurations exhibit a single transverse wave propagating towards the inner wall, accompanied by a pair of waves propagating towards the upper and lower walls. These wave patterns are consistent across varying mesh resolutions tested. Moreover, the velocities of propagation fall within the margin of measurement error. The computational cost ratio between the two mesh resolutions approximately follows $\left(\frac{\Delta h_{2}}{\Delta h_{1}}\right)^{4}$, indicating that refining the mesh by a factor of 2 requires 16 times computational resources. This highlights the need to balance resolution and computational expenses. For consistency, the cases examined in this study are conducted with a resolution of $\Delta h=1 / 20$.

\subsection{Effects of duct size (cases 11-14)}
\label{sec3.5}

In cases involving wide ducts ($W=20$), the simulation domain is expanded by a factor of $2 \times 2 \times 2$, also necessitating a significantly longer time to achieve statistical stability. Our analysis specifically considers rectangular perturbations, although the final propagation mode may not exclusively adhere to rectangularity. For most cases involving wide curved ducts, we analyze results after 1$\sim$10 cycles of detonation wave propagation around the circular path (where one cycle is defined as the detonation advancing $2 \pi$ angularly). Given the extensive simulation domain and prolonged physical time required, the wide duct cases are initially executed at a mesh resolution of $1 / 10$, subsequently refined to a finer resolution of $1 / 20$ for the final 40 time-units. Our focus here is primarily on both the straight duct case and the most curved $\left(R_{\mathrm{i}}=50\right)$ duct case.

\begin{figure}[t]
\centering
\includegraphics[width=15cm, trim=0cm 0cm 0cm 0cm, clip]{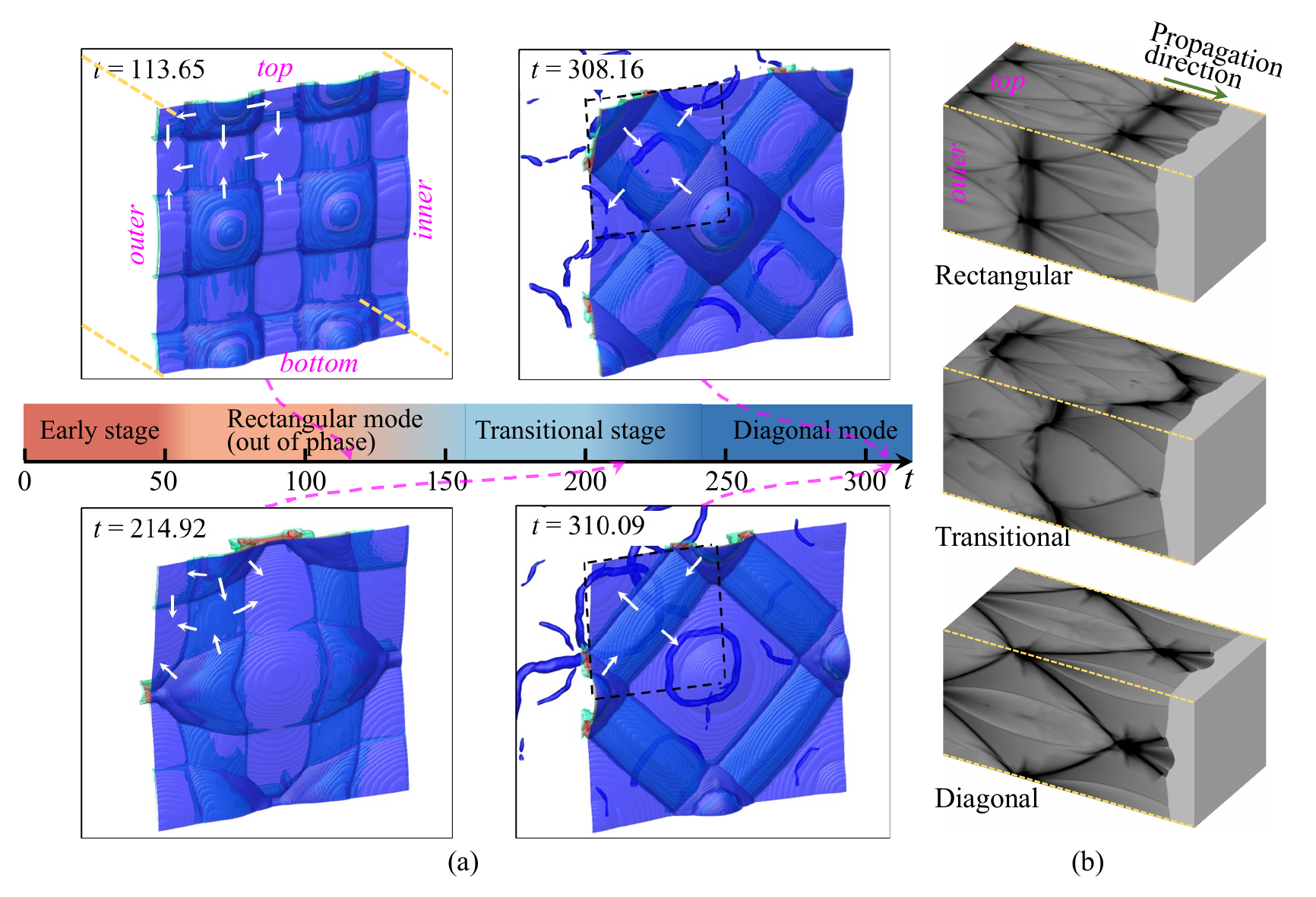}
\caption{Case 11: (a) Wave structures during mode transition in a wide straight duct. The structures in the black dashed zone are essentially the same to those in the narrow duct case with diagonal mode (see Fig. \ref{fig8}a). (b) Soot foils at different stages.}
\label{fig13}
\end{figure}

We first examine the straight duct case, Fig. \ref{fig13} provides an overview of the modes observed during propagation. Initially, a regular rectangular mode emerges after cellular structure formation, characterized by a noticeable half-phase shift between the cellular structures on adjacent walls (see soot foils in Fig. \ref{fig13}b). This can be observed in the transverse wave directions near the upper-left corner at $t=113.65$, where one wave moves towards the corner while another moves away from it. This behavior contrasts with the synchronized case observed at $t=72.15$, as illustrated in Fig. \ref{fig3}. Subsequently, transverse waves begin to distort and transition. In later stages, a diagonal mode becomes apparent. Interestingly, the wave structures within a quadrant of the wide domain closely resemble those observed in a narrower domain (Fig. \ref{fig8}). For instance, the wave patterns observed the upper-left quadrant at $t=308.16$ and $t=310.09$ closely resemble those at $t=56.5$ and $t=58.45$ in Fig. \ref{fig8}. It's noteworthy that a diagonal mode can manifest even without an initial diagonal perturbation. While there are two cells in the diagonal plane during diagonal mode, only one cellular structure without slapping wave is evident on the side walls, albeit rectangular and diagonal modes both containing four substructures. Therefore, caution is advised when correlating cell count with mode type and considering the presence of slapping waves. The conditions precipitating such transitions are currently constrained and warrant further investigation.

\begin{figure}[t]
\centering
\includegraphics[width=9cm, trim=0cm 0cm 0cm 0cm, clip]{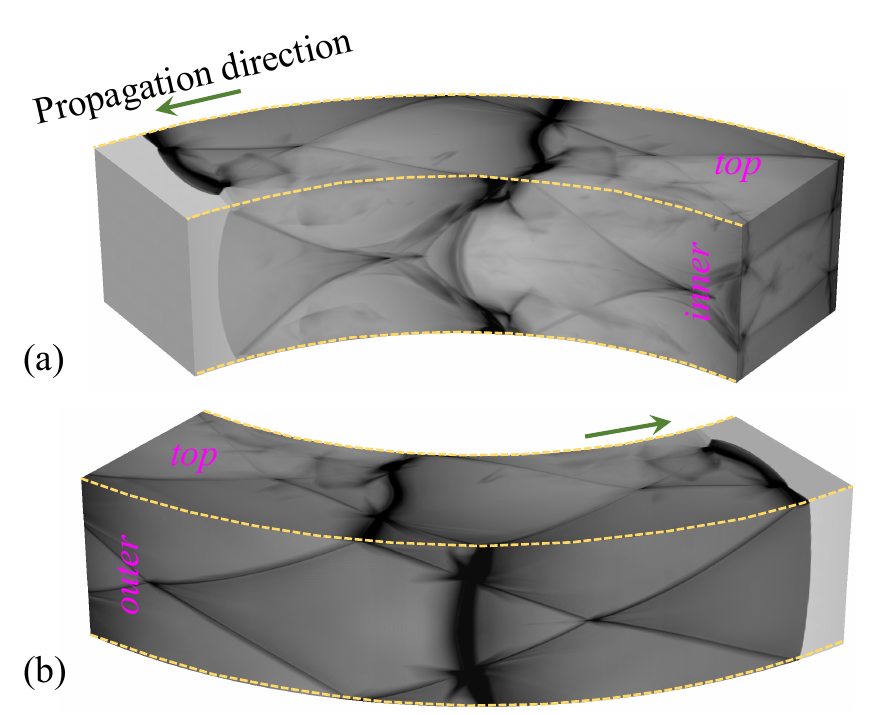}
\caption{Case 14: Soot foils at $t=226$ from two viewing angles for the wide duct case with $R_{\mathrm{i}} = 50$. Note that only two triple point trajectories are observed on each wall.}
\label{fig14}
\end{figure}

In the case of $R_{\mathrm{i}}=50$, the horizontal cell count remains consistently at one throughout the simulation. However, the number of cells on the two side walls is initially observed to be two, but this count decreases to a single cell after $t=131$, when the detonation wave has propagated approximately 2.3 rounds. The rectangular mode predominates during this period and remains consistent until the computation ends at $t=226$ (when the detonation wave has propagated approximately 4 rounds). This further suggests that detonation in a single bend may not accurately represent the fully developed detonation structures if the cross-section is large and the duct radius is small. Compared to the straight scenarios, whether in rectangular or diagonal modes, the duct's curvature significantly influences the cellular patterns on both the top/bottom and side walls (see Fig. \ref{fig14}).

As shown in Fig. \ref{fig10}, when the inner radius $R_{\mathrm{i}}$ is kept constant, increasing the width of the duct generally results in a decrease in speed along the inner wall. However, it is important to note that in sufficiently wide curved ducts, the speed may remain constant \citep{short2019propagation}. Nonetheless, conducting 3D simulations in ultra-wide curved ducts requires significant computational resources, which are beyond the scope of the present study.

\subsection{Effect of activation energy (cases 15-18)}
\begin{figure}[t]
\centering
\includegraphics[width=7cm, trim=0cm 0cm 0cm 0cm, clip]{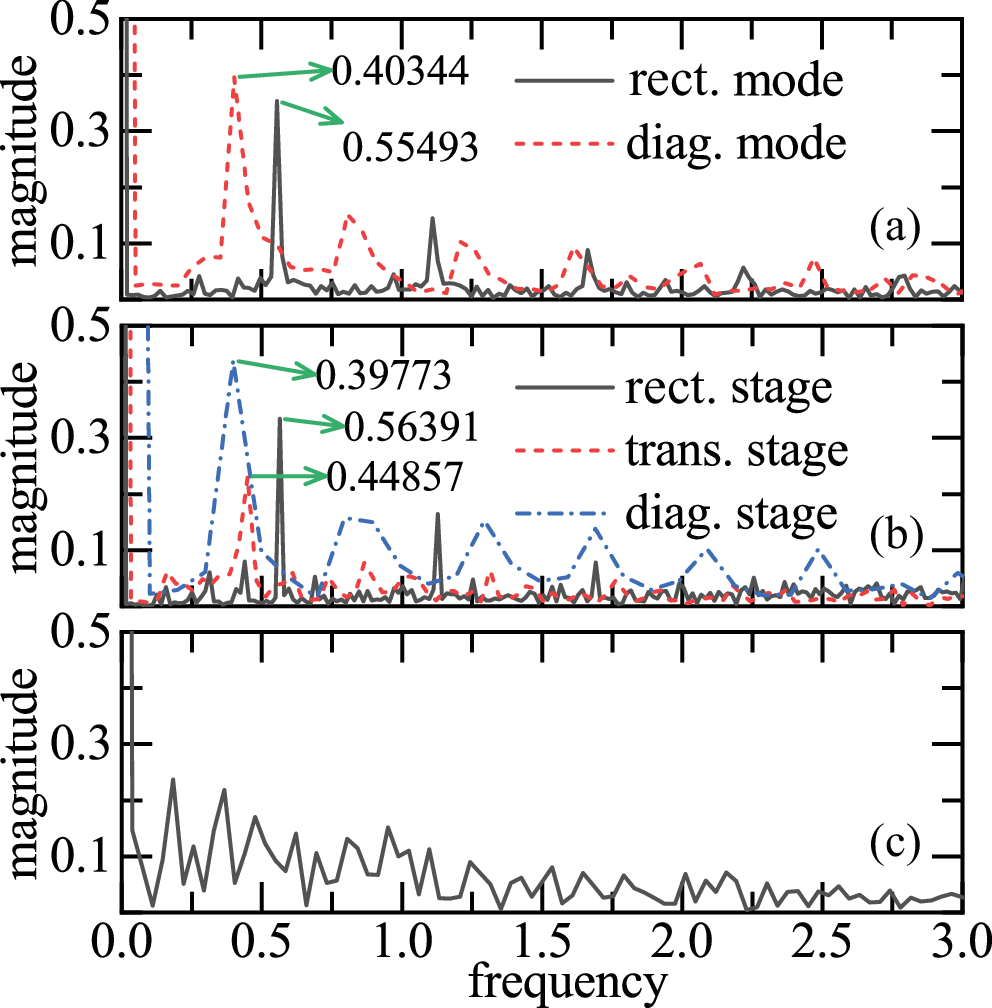}
\caption{FFT of velocities for (a) $E_{\mathrm{a}} = 10$, $W = 10$: Case 1 (rectangular mode) and Case 5 (diagonal mode); (b) Case 11: $E_{\mathrm{a}} = 10$, $W = 20$ showing rectangular, transitional, and diagonal stages; (c) Case 17: $E_{\mathrm{a}} = 50$, $W = 20$.}
\label{fig15}
\end{figure}

We proceed by investigating the impact of activation energy, specifically concentrating on cases with $E_{\mathrm{a}}=50$ within wide ducts. For high activation energy, the cellular structures exhibit significant irregularities, leading to substantial pressure fluctuations. Consequently, conducting statistical analysis across all four walls over a limited domain yields fewer uniform results compared to the scenario in straight ducts with $E_{\mathrm{a}}=10$. We first analyze the Fast Fourier Transform (FFT) of bulk velocity for both low and high activation energy scenarios in straight ducts (Fig. \ref{fig15}). For cases with regular cellular structures, the oscillation frequency is anticipated to correlate with the length of the cells and the velocity of the waves, thereby

$$
D \cdot \frac{1}{f}=\frac{L}{2}.
$$

\begin{figure}[t]
\centering
\includegraphics[width=15cm, trim=0cm 0cm 0cm 0cm, clip]{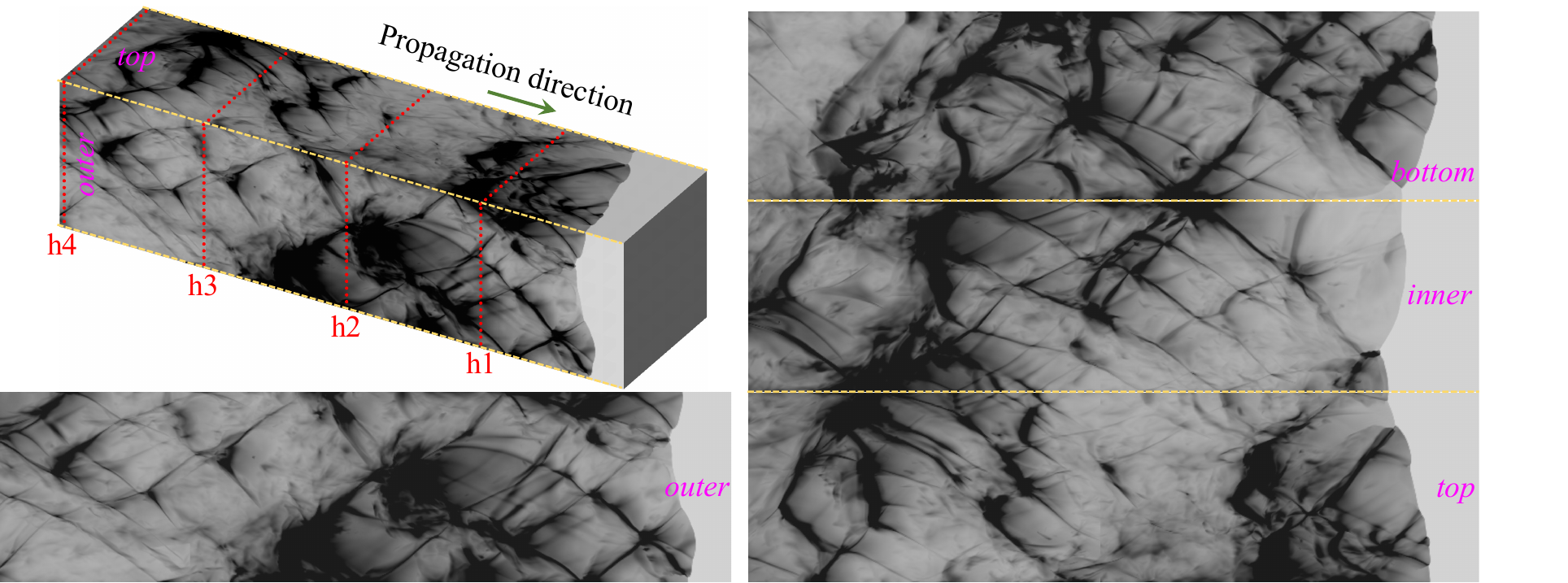}
\caption{Case 17: 3D view of soot foils alongside soot foils from each wall for a straight duct with $E_{\mathrm{a}} = 50$. The red dotted lines indicate the planes used to record the head-on soot foils.}
\label{fig16}
\end{figure}

Here, $D$ represents the average detonation wave velocity, $f$ denotes the oscillation frequency of the velocity, and $L$ signifies the length of the cell. The equation above suggests that during the propagation of a complete cell length, the velocity undergoes two identical profiles. For $E_{\mathrm{a}}=10$ in rectangular mode, the sampled cell length is approximately 24.4, corresponding to a theoretical frequency of $f=0.5582$. In diagonal mode under the same activation energy $E_{\mathrm{a}}=10$, the sampled cell length is about 32.6, indicating a frequency of $f=0.4178$. These values closely align with the peak frequencies obtained from the FFT analysis (Fig. \ref{fig15}a). The peak frequencies measured in the wide duct with the same $E_{\mathrm{a}}$ (Fig. \ref{fig15}b) are similar to those in the narrow duct cases. Conversely, for $E_{\mathrm{a}}=50$, there exists a broad spectrum where distinguishing a dominant frequency is challenging. The interaction of numerous transverse waves leads to significant pressure fluctuations and localized blasts. Observations from soot foil reveal extensive spatially non-uniform dynamics of detonation (Fig. \ref{fig16}).

\begin{figure}[t]
\centering
\includegraphics[width=15cm, trim=0cm 0cm 0cm 0cm, clip]{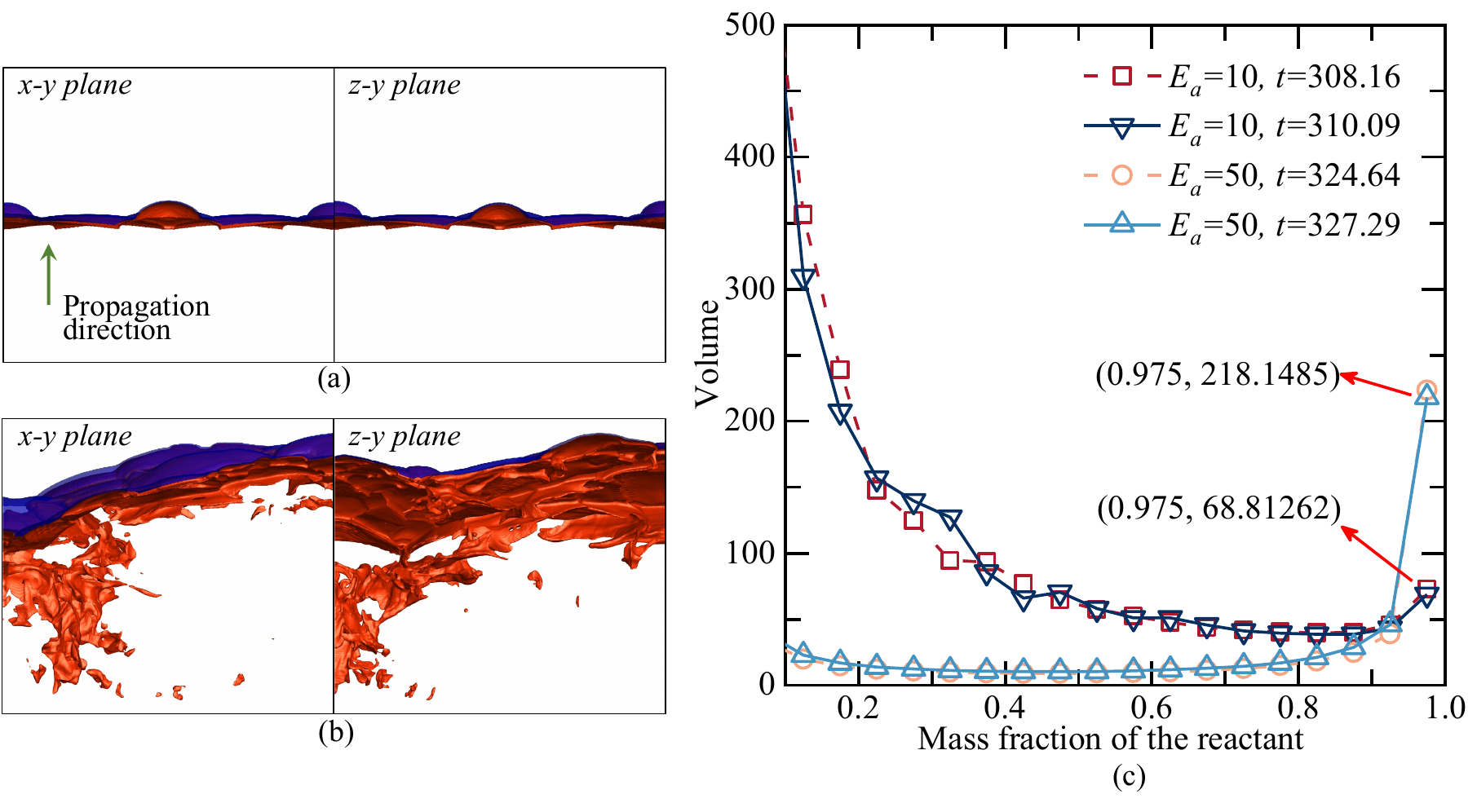}
\caption{Reactant distribution for (a) Case 11: $E_{\mathrm{a}} = 10$, $W = 20$, and (b) Case 17: $E_{\mathrm{a}} = 50$, $W = 20$. The blue contour indicates the shock front, while the red contour shows the iso-surface of $\lambda = 0.95$. (c) Histogram of gas with varying reactant mass fractions in the shocked region.}
\label{fig17}
\end{figure}

Additionally, Fig. \ref{fig17} illustrates the spatial distributions of the unreacted gas pockets and the volume of gas containing varying mass fractions of reactants behind the leading shock. It is evident that in the case of $E_{\mathrm{a}}=50$, a substantial volume of gas with $\lambda>0.95$ remains behind the shock.

\begin{figure}[t]
\centering
\includegraphics[width=15cm, trim=0cm 0cm 0cm 0cm, clip]{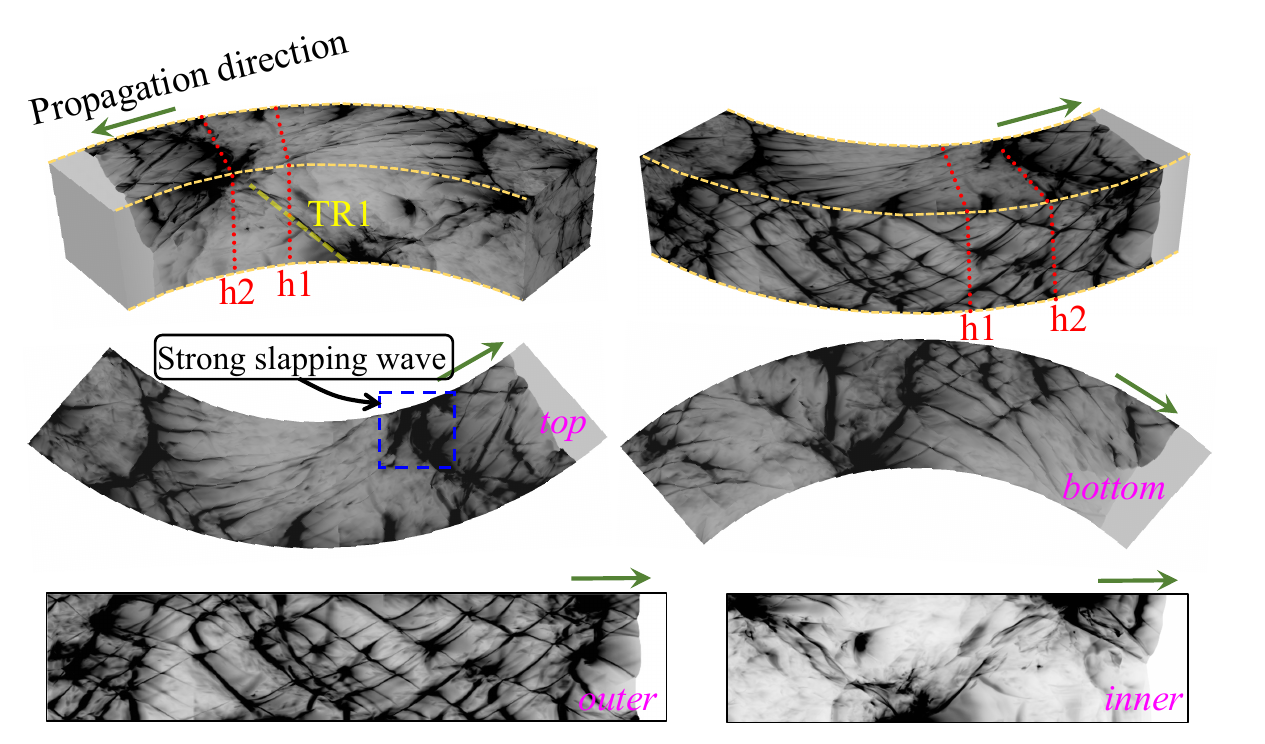}
\caption{Case 18: 3D view of soot foils from two angles, along with soot foils from each wall for a straight duct with $E_{\mathrm{a}} = 50$. TR1 represents the trajectory of the triple line near the inner wall. The red dotted lines indicate the planes used to record the head-on soot foils.}
\label{fig18}
\end{figure}

\begin{figure}[t]
\centering
\includegraphics[width=13cm, trim=0cm 0cm 0cm 0cm, clip]{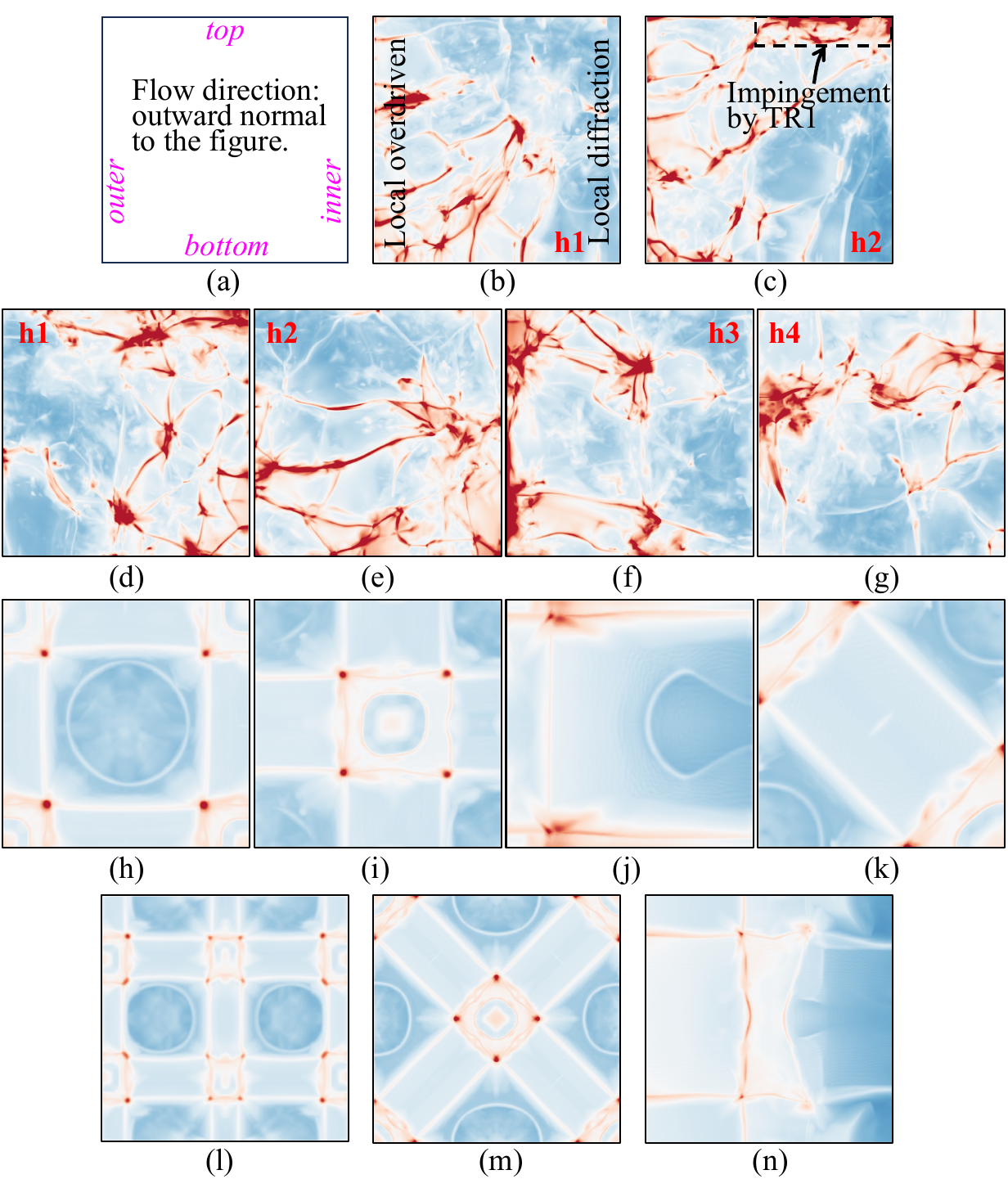}
\caption{Head-on soot foils: (a) Schematic; (b$\&$c) Case 18: $E_{\mathrm{a}} = 50$, $W = 20$, $R_{\mathrm{i}} = 50$; (d-g) Case 17: $E_{\mathrm{a}} = 50$, $W = 20$, $R_{\mathrm{i}} = \infty$; (h) Case 1: $E_{\mathrm{a}} = 10$, $W = 10$, $R_{\mathrm{i}} = \infty$; (i) Case 2: $E_{\mathrm{a}} = 10$, $W = 10$, $R_{\mathrm{i}} = 200$; (j) Case 4: $E_{\mathrm{a}} = 10$, $W = 10$, $R_{\mathrm{i}} = 50$; (k) Case 5: $E_{\mathrm{a}} = 10$, $W = 10$, $R_{\mathrm{i}} = \infty$ (diagonal mode); (l$\&$m) Case 11: $E_{\mathrm{a}} = 10$, $W = 20$, $R_{\mathrm{i}} = \infty$ (out-of-phase rectangular mode and diagonal mode); (n) Case 14: $E_{\mathrm{a}} = 10$, $W = 20$, $R_{\mathrm{i}} = 50$.}
\label{fig19}
\end{figure}

In the straight duct cases with high $E_{\mathrm{a}}$ mentioned above, despite the absence of clearly discernible structured patterns, the random structures on all four walls appear indistinguishable. However, this uniformity changes significantly in the scenario with $R_{\mathrm{i}}=$ 50 (Fig. \ref{fig18}). Distinct patterns emerge: the outer wall exhibits smaller cell sizes, contrasting sharply with the inner wall, where a single transverse wave oscillates back and forth. On the top and bottom walls, cell patterns are concentrated near the outer edge, accompanied by pronounced slapping impacts from the vigorous wave motion observed on the inner wall. This behavior arises because the reaction rate for gases with high activation energy is highly sensitive to local velocity deficits near the inner wall and the overdriven state near the outer wall. The high activation energy, combined with gas expansion near the inner wall, leads to localized decoupling of the shock and flame. This region of gas subsequently interacts with transverse waves propagating horizontally and strong slapping waves moving vertically (e.g., TR1), both clustered near the inner wall. This can be further illustrated by examining the head-on soot foils, with locations indicated by red dotted lines in Fig. \ref{fig18}. At location h1, multiple transverse waves are clustered near the outer region, while pressure near the inner part remains relatively low (Fig. \ref{fig19}b). Additionally, Fig. \ref{fig19}c highlights the strong impingement of the transverse wave TR1. In contrast, the head-on soot foils in the straight case (Fig. \ref{fig19}d-g) exhibit randomly distributed fine cellular structures. Furthermore, Figs. \ref{fig19}h-n present selected representative detonation structures for cases with low activation energy, varying cross-sectional sizes, and inner radii, and these figures are self-explanatory.

\begin{figure}[t]
\centering
\includegraphics[width=14cm, trim=0cm 6cm 0cm 0cm, clip]{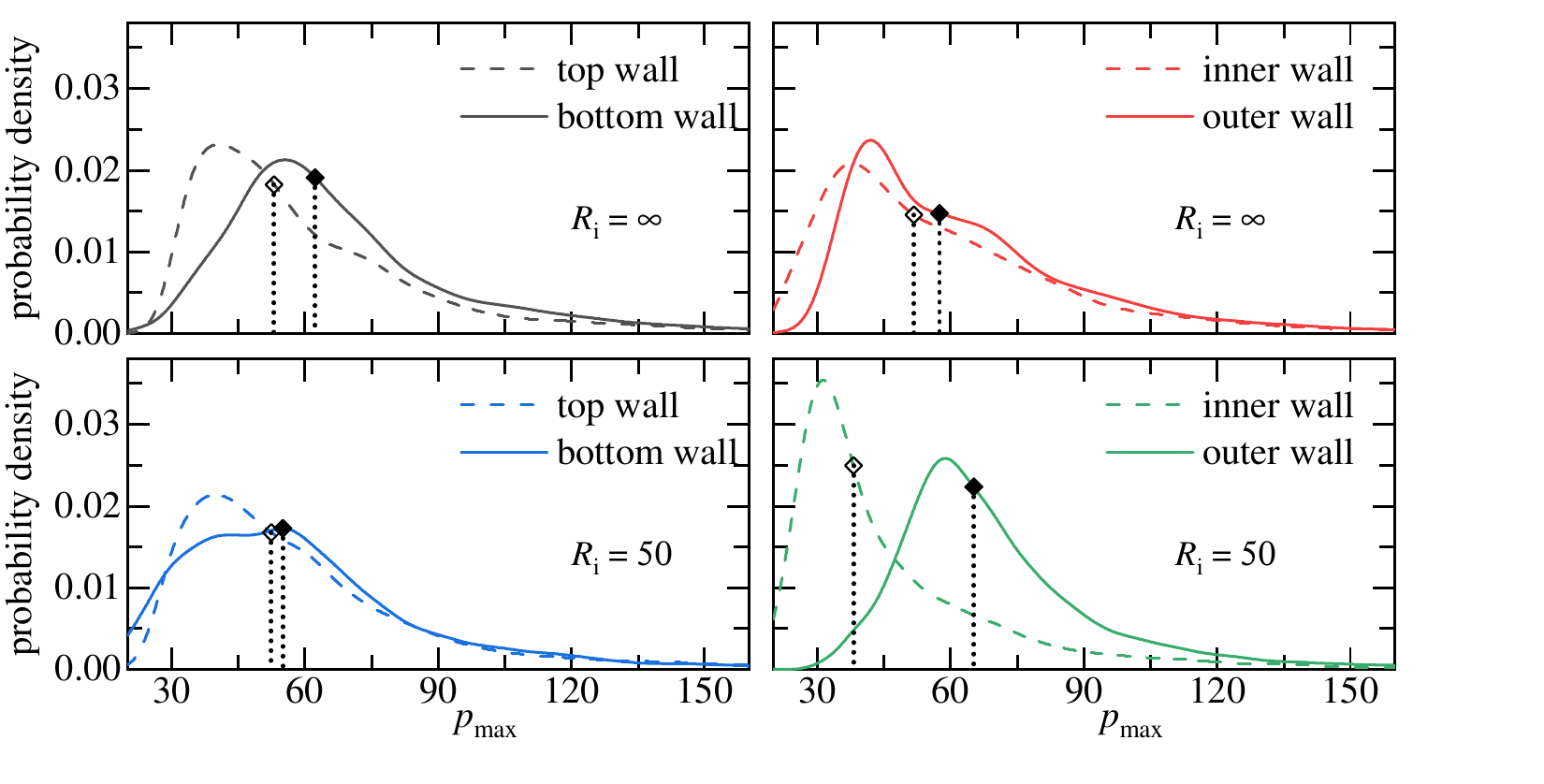}
\caption{Histograms of maximum pressure recorded on all walls for Case 17: $R_{\mathrm{i}} = \infty$ and Case 18: $R_{\mathrm{i}} = 50$. Symbols with drop lines indicate the median values.}
\label{fig20}
\end{figure}

Figure \ref{fig20} depicts histograms for both straight and curved cases with high activation energy. It is noteworthy that unlike the scenario with low activation energy, where wave patterns and histograms exhibit nearly symmetric distributions for straight duct cases, the current histograms may vary depending on the sampling domain. Figure \ref{fig20} is confined to the domain illustrated in Figure \ref{fig16}, where the maximum pressure unevenly distributes across the four walls within this limited area. Furthermore, in comparison to Fig. \ref{fig7}, the values are less concentrated, and the maximum probability density is notably lower. In the case of the curved duct, the median value on the inner wall is significantly smaller than that on the outer wall, primarily due to the substantial amounts of unreacted gas that are ultimately consumed by the strong transverse wave. Conversely, the outer wall displays numerous small cellular structures, leading to a shift of maximum pressure towards higher values. The top and bottom walls exhibit a mixture of high-pressure regions characterized by cellular structures and slapping waves, along with low-pressure areas, resulting in histograms that appear more flattened compared to those of the side walls.

\begin{figure}[t]
\centering
\includegraphics[width=7cm, trim=0cm 0cm 0cm 0cm, clip]{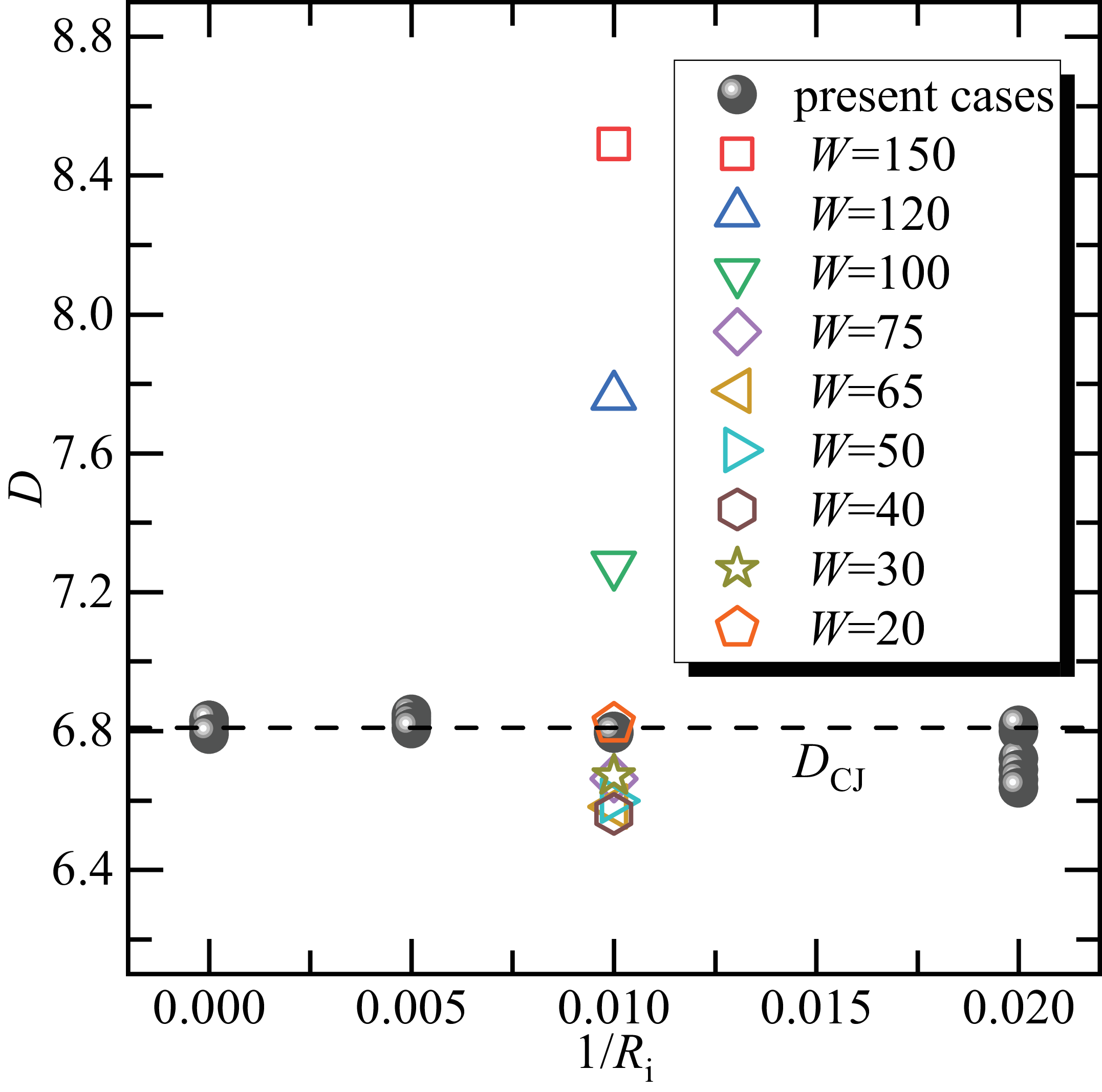}
\caption{Average detonation velocity along the centerline of the duct for all cases in the present study (at $r = R_{\mathrm{i}} + \frac{W}{2}$). The colored symbols represent the 2D results from \citet{short2019propagation}.}
\label{fig21}
\end{figure}

The velocities shown in Fig. \ref{fig10} are measured over a substantial distance of more than 130 length units. Interestingly, it is observed that the velocities in the cases of this study depend only on $W$ and $R_{\mathrm{i}}$. Whether in 2D or 3D, and across different $E_{\mathrm{a}}$ values and various propagation modes, these factors do not significantly influence propagation velocity, despite potential differences in wave structures and statistical distributions. This phenomenon is reminiscent of classical 1D detonation theory, where $D \sim \sqrt{Q}$ \citep{lee2008detonation}. However, the theoretical foundation for continuous curved ducts remains inadequately defined. Furthermore, Fig. \ref{fig21} illustrates the average velocities along the centerline of the duct (in the circular direction at $r=R_{\mathrm{i}}+W / 2$). The velocities along the centerline roughly approximate the CJ value, particularly for larger $R_{\mathrm{i}}$. It is important to recognize that velocities measured along the centerline may not essentially align with the CJ solution. As discussed in previous 2D research \citep{short2019propagation}, in sufficiently wide curved channels, the angular velocity remains constant due to the presence of a ``detonation driving zone'' structure. However, this aspect is currently beyond the capabilities of the present 3D simulations. Consequently, in such configurations, centerline velocities may diverge significantly from the CJ value.

\section{Conclusions}
In this study, we have investigated three-dimensional detonation waves within continuously curved ducts with square cross-section, presenting findings based on wave structures, soot foil analysis, velocity evolutions, and pressure distribution analyses. For cases with narrow ducts, the comparisons with straight duct configurations reveal notable variations in propagation modes. Specifically, as inner radius decreases, we observe a slower detonation velocity along the inner wall, a transition in cellular structure from regular to skewed, and a reduction in the number of transverse waves on the top and bottom walls, shifting from two waves to one for sufficiently small inner radii. Using a diagonal perturbation, the diagonal mode is observed for straight and large inner radius ducts, for small ones, only rectangular mode is observed due to the compression from the interactions with the side walls. Additionally, simulations involving wide ducts derived from narrow cases show the emergence of four regular cells and a transition from out-of-phase rectangular mode to diagonal mode in straight duct configurations. Despite similarities in velocity deficits and reductions in transverse waves, stability in wide ducts requires significantly longer distances. Finally, the effects of activation energy are considered. The analysis of wave structures and pressure distributions reveals smaller cellular structures near the outer wall, while larger wave motions occur near the inner wall in the case of a small inner radius. Overall, our findings underscore the inherent complexity of three-dimensional detonation waves in curved ducts, highlighting substantial qualitative and quantitative differences compared to straight ducts. The design and evaluation of practical rotating detonation engines should take into account the three-dimensional effects associated with duct curvature, rather than relying solely on insights from two-dimensional or straight three-dimensional detonation dynamics.

\clearpage

\appendix
\renewcommand{\thefigure}{A\arabic{figure}} 
\setcounter{figure}{0} 
\section*{Appendix}
\label{appen}
\section*{A1. Validation for detonation in a 2D curved channel}
\begin{figure}[H]
\centering
\includegraphics[width=11.5cm, trim=1.2cm 0cm 1.2cm 9cm, clip]{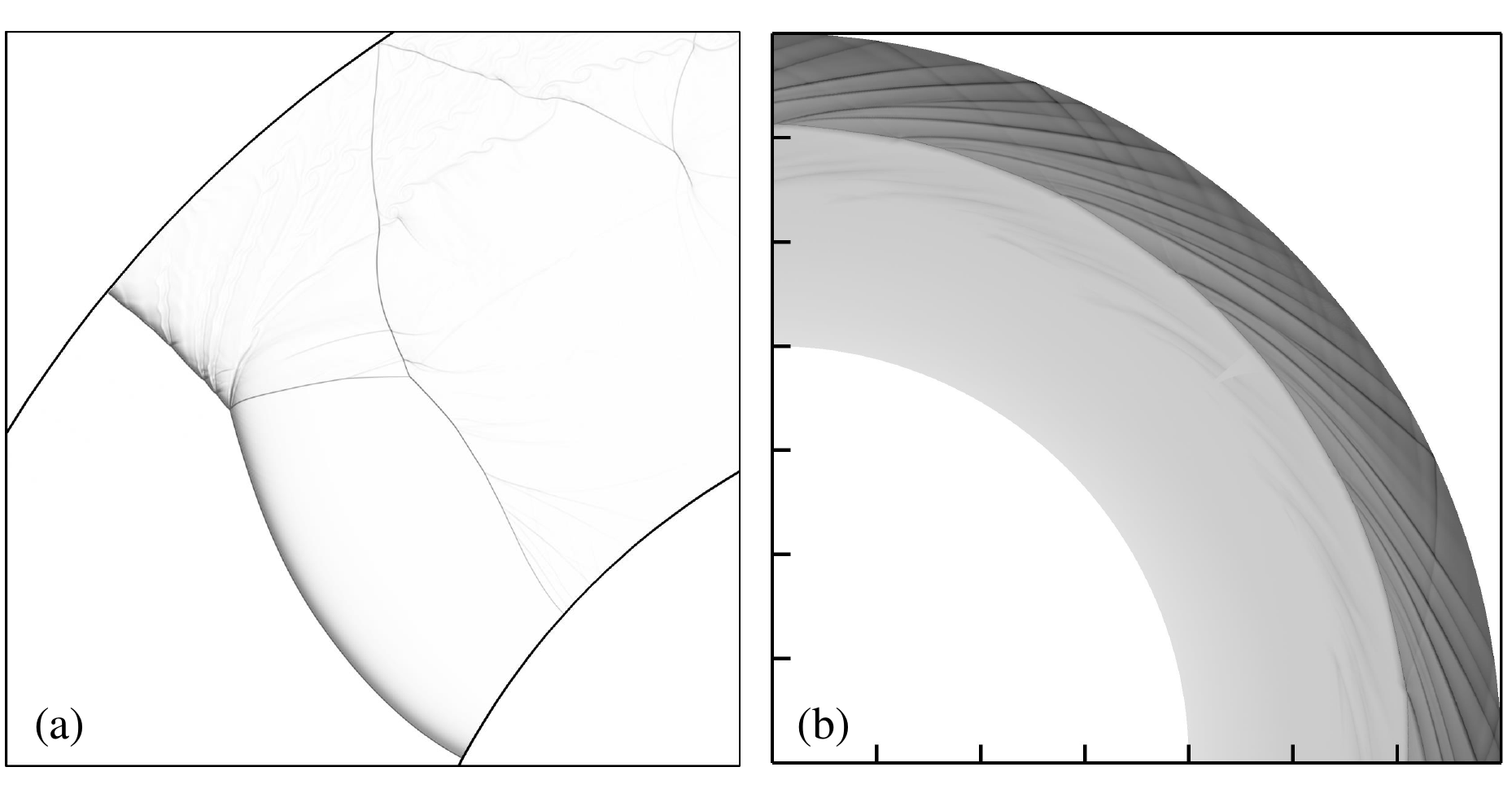}
\caption{Detonation in a wide curved channel. (a) Numerical schlieren image; (b) Numerical soot foils.}
\label{A1}
\end{figure}

In \citet{short2019propagation}, the flow equations are integrated using second-order schemes on 2D Cartesian meshes, employing a block-structured adaptive mesh refinement method to achieve a resolution of 20 points per half-reaction length. The boundary conditions for the arc surfaces are implemented using an embedded internal boundary technique. In the present study, we utilize grid transformation techniques to create structured curvilinear meshes that conform precisely to the arc boundaries. We maintain the same mesh resolution and employ a moving domain technique. The 2D computation is conducted by setting the length along the $z$-direction to be less than one, simulating 2D scenarios. The parameters selected are $E_{\mathrm{a}}=10$, $Q=50$, $\gamma=1.2$, $R_{\mathrm{i}}=100$, and $R_{0}=175$. Initially, a perturbed ZND detonation is implemented. Figure \ref{A1} presents the schlieren image at the detonation wave front along with the corresponding numerical soot foil. The fully developed detonation wave exhibits a curved and smooth front, with a Mach stem located near the outer wall. The computed angular velocity of the detonation wave near the inner wall is 0.0483 , which aligns closely with the result from \citet{short2019propagation} (approximately 0.04845 ). This test validates the algorithm's effectiveness in computing detonation propagation within a continuously 2D curved channel.

\section*{A2. Validation for 3D detonation in a straight tube}
\begin{figure}[H]
\centering
\includegraphics[width=14.5cm, trim=3cm 3cm 1cm 0cm, clip]{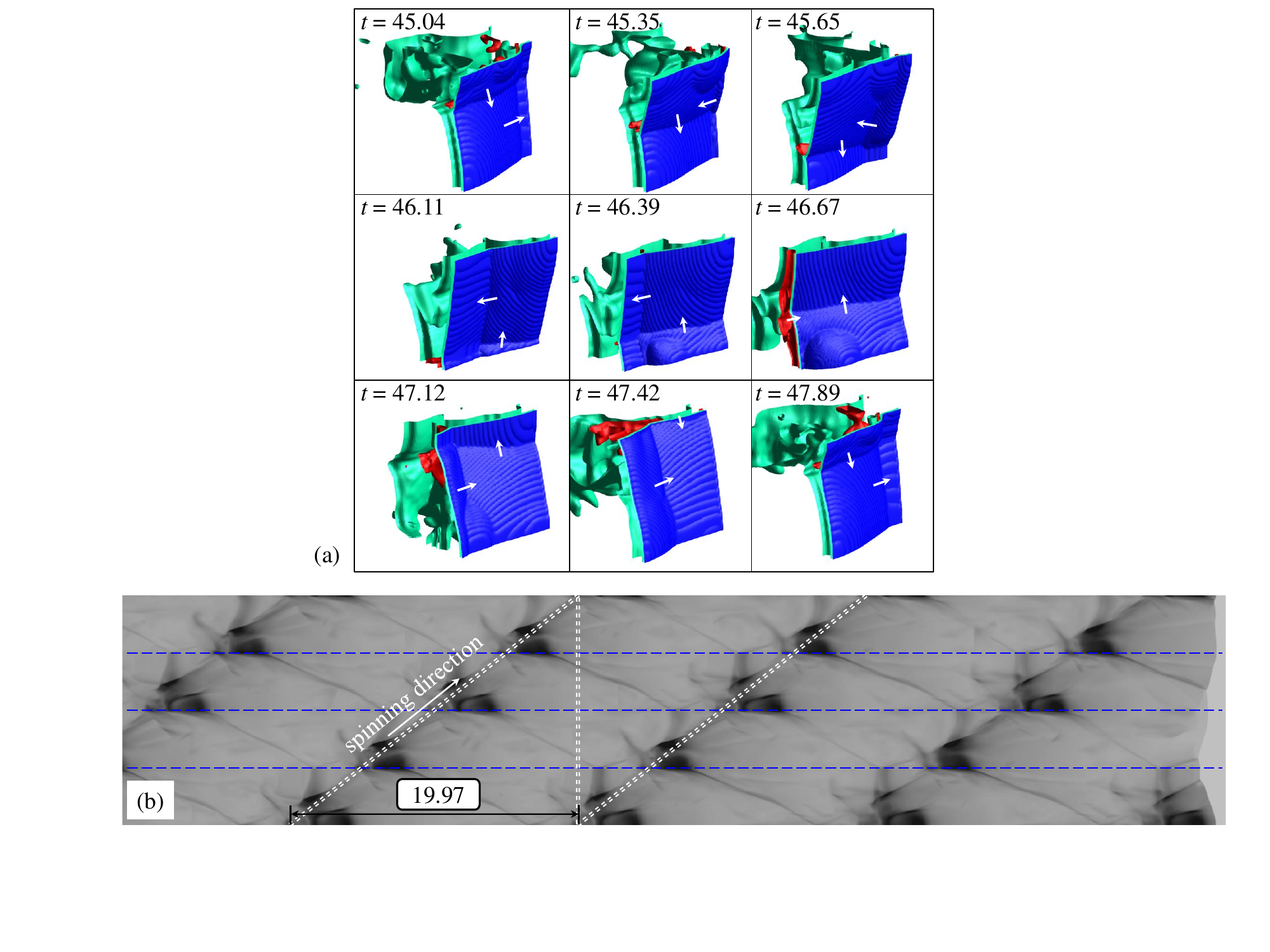}
\caption{Spinning detonation in a narrow straight duct. (a) Density iso-contours within one spinning cycle; (b) Numerical soot foils on each wall.}
\label{A2}
\end{figure}

In \citet{dou2008simulations}, high-resolution simulations were conducted to study 3D detonation wave propagation in a straight duct. In this study, we focus on a narrow duct case which features a square cross-section of $4 \times 4$, with parameters $E_{\mathrm{a}}=20$, $Q=50$, and $\gamma=1.2$. \citet{dou2008simulations} observed a spinning motion in this configuration. Figure \ref{A2} illustrates the wave structures and numerical soot foil simulated by the current solver. A spinning detonation wave is clearly formed, and the distance between successive pressure peaks aligns well with the reference findings. This test confirms the reliability of the solver in accurately computing 3D detonation waves in a straight duct.

Considering the results from tests A1 and A2, along with convergence tests from Sec. 3.4, the current solver is deemed reliable and accurate.

\bibliographystyle{apalike}
\bibliography{curved3DBib} 

\end{document}